%% file: BarHalo_arXiv2nd.tex
\documentclass[fleqn,usenatbib]{mnras}

\usepackage{newtxtext,newtxmath}

\usepackage[T1]{fontenc}

\DeclareRobustCommand{\VAN}[3]{#2}
\let\VANthebibliography\thebibliography
\def\thebibliography{\DeclareRobustCommand{\VAN}[3]{##3}\VANthebibliography}

\usepackage{graphicx}	
\usepackage{amsmath}	
\usepackage{amssymb}	
\usepackage{bm}
\usepackage{caption}
\usepackage{multicol}
\usepackage{hyperref}
\usepackage{booktabs}
\usepackage{ulem}

\input{./newcommand}

\title[]{Oscillating dynamical friction on galactic bars by trapped dark matter}

\author[R. Chiba et al.]{
Rimpei Chiba$^{1}$\thanks{E-mail: rimpei.chiba@physics.ox.ac.uk}
and Ralph Sch{\"o}nrich$^{2}$
\\
$^{1}$Rudolf Peierls Centre for Theoretical Physics, Clarendon Laboratory, Parks Road, Oxford OX1 3PU, UK\\
$^{2}$Mullard Space Science Laboratory, University College London, Holmbury St. Mary, Dorking, Surrey, RH5 6NT, UK
}

\date{Accepted XXX. Received YYY; in original form ZZZ}

\pubyear{2021}

\begin{document}
\label{firstpage}
\pagerange{\pageref{firstpage}--\pageref{lastpage}}
\maketitle


\begin{abstract}

The dynamic evolution of galactic bars in standard $\Lambda$CDM models is dominated by angular momentum loss to the dark matter haloes via dynamical friction. Traditional approximations to dynamical friction are formulated using the linearized collisionless Boltzmann equation and have been shown to be valid in the \textit{fast limit}, i.e. for rapidly slowing bars. However, the linear assumption breaks down within a few dynamical periods for typical slowly evolving bars, which trap a significant amount of disc stars and dark matter in resonances. Recent observations of the Galactic bar imply this \textit{slow regime} at the main bar resonances. We formulate the time-dependent dynamical friction in the \textit{slow limit} and explore its mechanism in the general slow regime with test-particle simulations. Here, angular momentum exchange is dominated by resonantly trapped orbits which slowly librate around the resonances. In typical equilibrium haloes, the initial phase-space density within the trapped zone is higher at lower angular momentum. Since the libration frequency falls towards the separatrix, this density contrast winds up into a phase-space spiral, resulting in a dynamical friction that oscillates with $\sim$Gyr periods and damps over secular timescales. We quantify the long-term behaviour of this torque with secular perturbation theory, and predict two observable consequences: i) The phase-space spirals may be detectable in the stellar disc where the number of windings encodes the age of the bar. ii) The torque causes weak oscillations in the bar's pattern speed, overlaying the overall slowdown -- while not discussed, this feature is visible in previous simulations.

\end{abstract}




\defcitealias{lynden1972generating}{LBK72}
\defcitealias{Tremaine1984Dynamical}{TW84}
\defcitealias{Weinberg2004Timedependent}{W04}

\section{Introduction}
\label{sec:introduction}

Dark matter haloes strongly affect the evolution of galactic bars: dynamical friction transfers angular momentum and energy from the galactic bar to the dark matter halo, allowing bars to slow and grow. This angular momentum transfer depends both on the density and on the kinematic state of the dark halo. Consequently, we can constrain the phase-space distribution and nature of dark haloes from the measured slowing rate of bars.

While a wide range of numerical simulations have addressed the angular momentum transfer from bars to haloes \citep[e.g.][]{hernquist1992bar,debattista2000constraints,athanassoula2003determines}, the analytical treatment is still incomplete. \citet[][hereafter \citetalias{Tremaine1984Dynamical}]{Tremaine1984Dynamical} derived a formula of dynamical friction in an inhomogeneous spherical system using linear perturbation theory. Their formula generalises the earlier torque formula by \citet[][hereafter \citetalias{lynden1972generating}]{lynden1972generating}, which describes the angular momentum transport in two-dimensional disc galaxies. Unlike the dynamical friction in a homogeneous system \citep{Chandrasekhar1943DynamicalFriction}, the LBK formula shows that in an inhomogeneous system, where orbits are quasi-periodic, the dominant angular momentum transfer arises from discrete resonances. \citetalias{Tremaine1984Dynamical} also investigated the non-linear motion of orbits near resonances and showed that the LBK formula is only valid in what they call the \textit{fast limit}. Here, `fast' means that the resonances sweep across the phase space of the halo rapidly enough to preclude the growth of non-linear responses, in particular resonant trapping. This allows the system to remain in the linear regime. \cite{weinberg1985evolution} showed that the LBK torque formula predicts a too strong friction on the bar, extracting most of its angular momentum in a few rotation periods. This problem was revisited by \citet[][hereafter \citetalias{Weinberg2004Timedependent}]{Weinberg2004Timedependent}, where he pointed out that the assumption of the perturbation being turned on adiabatically in the distant past (the time-asymptotic limit) is problematic since the evolution timescale of the bar is only several times the characteristic dynamical period. Relaxing this assumption enabled \citetalias{Weinberg2004Timedependent} to model the transient effect and reduce the overall torque. Yet, the linear treatment by \citetalias{Weinberg2004Timedependent} remains restricted to the fast regime and cannot be applied to the \textit{slow regime} where resonant trapping comes into play.

This slow regime (resonant trapping by the Galactic bar) is, however, implied by the stellar kinematics of the Solar neighbourhood observed by the \textit{Gaia} satellite \citep[e.g.][]{GaiaDR2Katz2019,Hunt2019signature,Monari2019signatures,Binney2020Trapped,Trick2021Identifying}. The resonantly trapped stars manifest as stellar streams which occupy a sizeable fraction of the local phase-space volume. Modelling particularly the Hercules stream, \cite{Chiba2020ResonanceSweeping} inferred that the bar is decelerating at a moderate rate, keeping a good portion of local stars trapped. Furthermore, signatures of trapping is found in the metallicity distribution of local stars: the corotation resonance of the slow/long bar exhibits a monotonic increase in metallicity towards the core of the resonance, indicating a tree-ring like growth as predicted by a naturally slowing bar that sequentially captured stars from small Galactocentric radii \citep{Chiba2021TreeRing}.

Resonant trapping by the bar is also commonly seen in self-consistent \textit{N}-body simulations: spectral analysis of orbits in barred galaxies shows significant clustering near resonances \citep[e.g.][]{Ceverino2007Resonances}. Simulations by \cite{Halle2018Radial} show that these trapped stars adhere to the resonance as the bar slows down. Some works have also discussed direct trapping of dark matter into a `shadow bar' \citep{Athanassoula2007BarInHalo,Petersen2016DarkMatterTrapping,Collier2021Coupling}.

Thus, both observations and simulations indicate that the main resonances of typical bars evolve in the slow regime. Generally, bar evolution involves both fast and slow regimes depending on the evolutionary phase, the type of resonance, and the position along each resonance, since fast or slow is decided by the measure of the bar's slowing rate against the local libration frequency. While the dynamical friction in the fast limit has been modelled successfully by linear theory (\citetalias{Weinberg2004Timedependent}), the slow regime has been little explored. The sole exception is \citetalias{Tremaine1984Dynamical} who formulated the net change in angular momentum of orbits swept past by a resonance by analytically integrating the torque from $t=-\infty$ to $\infty$. Their formula, however, does not describe the dynamical friction due to the transient response against the passage of the resonance (e.g. phase mixing). Similarly to the point raised by \citetalias{Weinberg2004Timedependent} in the fast limit, we show that the transient effects are significant and even more so in the slow regime due to the long mixing time of trapped orbits. Therefore, beyond the work of \citetalias{Tremaine1984Dynamical}, modelling bar evolution requires a time-dependent theory of dynamical friction in the slow regime. This paper takes the first step towards this problem by working in the \textit{slow limit} of bar evolution (i.e. no slowdown). Taking the slow limit enables us to analytically model dynamical friction by both trapped and untrapped orbits in a fully time-dependent manner while setting aside the more complicated process of resonant capture and escape \citep[e.g.][]{Henrard1982Capture,Sridhar1996Adiabatic}. We will, however, show numerically that the qualitative mechanism of dynamical friction identified in the slow limit applies generally to the slow regime.

The dynamics of resonant trapping is best viewed in the slow angle-action plane which exhibits a phase flow akin to that of a simple pendulum: trapped orbits librate around the resonance centre (and their slow angle oscillates around the resonance midpoint), while untrapped orbits circulate above and below the separatrix (i.e. their slow angle moves freely across the whole $2\pi$ range). Before bar formation, the phase-space density typically declines towards large slow action (i.e. the $z$-angular momentum). When the bar forms, the trapped region grows and captures orbits from both above and below the resonance, reconnecting phase space across regions of different densities. The newly trapped orbits are thus inevitably non-uniform in the libration angle. As these trapped orbits librate around the resonance along contours of constant libration actions, their innate angle imbalance results in a periodic dynamical friction on the bar. However, due to the monotonic decrease of libration frequency towards the separatrix, the density inside the trapped zone gradually winds up into a phase-space spiral, and the net torque from trapped orbits slowly attenuates. The phase mixing timescale is set by the libration period (typically $> 1 \Gyr$), which prevents complete mixing in a Hubble time. This leads us to predict that bars are subject to pattern speed oscillations on the timescale of the libration, which is an order of magnitude longer than the short-period fluctuations due to bar-spiral interactions \citep{Wu2016TimeDependent,Hilmi2020Fluctuations}.

In this paper, we study this behaviour both analytically and in test particle simulations. The analytical approach uses the resonant angle-action coordinates for each resonance obtained by the method of averaging \citep[e.g.][]{lichtenberg1992regular}. These local coordinates allow us to model the resonant dynamics trivially. We also discuss the accuracy of the averaging method when chaos emerges due to the overlap of multiple resonances.

This paper is organized as follows. In section \ref{sec:model}, we introduce our model of the halo and the bar. Section \ref{sec:torque_fast_regime} reviews the torque formula in the fast limit and discusses its problems when applied to systems in the slow regime. In section \ref{sec:torque_slow_regime}, we lay out the standard method to treat the non-linear dynamics near resonance and formulate the dynamical friction in the slow limit which we verify against test-particle simulations. We will discuss the consequences of an evolving bar in section \ref{sec:slowdown_bar} and sum up in section \ref{sec:conclusions}.

\section{Model}
\label{sec:model}

\subsection{Model of dark halo}
\label{sec:dark_halo}

We model the dark halo with an isotropic \cite{Hernquist1990AnalyticalModel} model:
\begin{align}
  \rho_0(r) &= \frac{M}{2\pi} \frac{\rs}{r(\rs+r)^3},
  \label{eq:Hernquist_density} \\
  \Phi_0(r) &= - \frac{GM}{\rs+r},
  \label{eq:Hernquist_potential}
\end{align}
where $G$ is the gravitational constant, $M$ is the total halo mass, and $\rs$ the halo scale radius. By default, we set $M = 1.5 \times 10^{12} \Msun$ and $\rs = 20 \kpc$. The corresponding distribution function and setup of the test-particle simulations are described in Appendix \ref{sec:app_DF}.

\subsection{Model of Galactic bar}
\label{sec:bar}

We model the bar with a quadrupole rotating at pattern speed $\Omegap$:
\begin{align}
  \Phib(r,\vartheta,\varphi,t) = \Phib(r) \sin^2 \vartheta \cos 2 \left(\varphi - \Omegap t\right),
  \label{eq:bar_potential}
\end{align}
where $(r,\vartheta,\varphi)$ are the usual spherical coordinates. The radial dependence of the bar potential is \citep{Chiba2020ResonanceSweeping}
\begin{align}
  \Phib(r) = - \frac{A \vc^2}{2} \left(\frac{r}{\rCR}\right)^2 \left(\frac{b + 1}{b + r/\rCR}\right)^5,
  \label{eq:bar_potential_amp}
\end{align}
where $A$ is the dimensionless strength of the bar, $\vc$ is the local circular velocity, and $b$ describes the ratio between the bar scale length and the corotation radius $\rCR$. This potential is designed to lengthen as the bar slows.   
We fit our model to that of \cite{Sormani2015GasIII} at the Galactic plane ($\vartheta = \pi/2$) and obtain $A=0.02, b=0.28$. By default, we set $\vc = 235 \kms$ and $\Omegap = 35 \kmskpc \simeq 35.8 \iGyr$ \citep{Binney2020Trapped,Chiba2021TreeRing,Clarke2021ViracGaia}.

\subsection{Angle-action variables in spherical potential}
\label{sec:coordinate}

The description of orbits in an integrable system is greatly simplified with the use of angle-action variables $(\vtheta,\vJ)$ \citep[e.g.][]{binney2008galactic}. By construction, the Hamiltonian $H_0(\vJ)$ does not depend on the angles, so the equations of motion are trivial in these coordinates:
\begin{align}
  \dot{\vJ} = - \frac{\pd H_0}{\pd \vtheta} = {\bm 0}, ~~~~~~
  \dot{\vtheta} = \frac{\pd H_0}{\pd \vJ} \equiv \vOmega(\vJ),
  \label{eq:dJdt}
\end{align}
i.e., the actions are conserved and the conjugate angles increase linearly with time at constant rate $\vOmega(\vJ)$. 

In a spherically symmetric potential, a convenient set of actions are $\vJ=(\Jr,L,\Lz)$ where $\Jr$ is the radial action, $L$ is the magnitude of the angular momentum vector ${\bm L}$, and $\Lz$ is the $z$-component of ${\bm L}$. The ratio between $L$ and $\Lz$ defines the orbital inclination $\beta \equiv \cos^{-1}\left(\Lz/L\right)$. The conjugate angle variables $\vtheta=(\thetar,\thetapsi,\thetaphi)$ describe, respectively, the radial phase, the azimuthal phase in the orbital plane, and the (fixed) longitude of the ascending node, i.e. the intersection of the orbital plane and the galactic equatorial plane.

\section{Torque in the fast limit}
\label{sec:torque_fast_regime}

To compare our approach (section \ref{sec:torque_slow_regime}) with the conventional approximations, we here briefly review the torque formula in the fast limit developed by \citetalias{Tremaine1984Dynamical} and \citetalias{Weinberg2004Timedependent}. Following \citetalias{Weinberg2004Timedependent}, we solve for the second-order torque using the linear solution to the collisionless Boltzmann equation. We will demonstrate the breakdown of the formula in the slow regime and discuss the cause of the problem. Throughout the paper, we ignore the halo's self-gravitational perturbation. 

We assume that the halo is in a steady state prior to bar formation. The unperturbed distribution is then a function of the actions alone $f_0(\vJ)$ (Jeans' theorem). When the bar forms, the perturbing potential changes the halo's distribution by $f_1(\vtheta, \vJ, t)$. We can then write the Hamiltonian $H$ and the distribution function $f$ as
\begin{align}
  H(\vtheta, \vJ, t) = H_0(\vJ) + \Phi_1(\vtheta, \vJ, t), ~ 
  f(\vtheta, \vJ, t) = f_0(\vJ) + f_1(\vtheta, \vJ, t),
  \label{eq:H0_H1_f0_f1}
\end{align}
where $\Phi_1$ consists of the bar potential (\ref{eq:bar_potential}) and the self-gravitational perturbation induced by the response of the halo $f_1$, although here we ignore the latter. The angle variables are, by construction, $2\pi$ periodic, so one can expand the perturbations into a Fourier series:
\begin{align}
  \Phi_1(\vtheta, \vJ, t) = \sum_{\vn} \hPhin(\vJ, t) \e^{i \vn \cdot \vtheta}, ~~~ 
  f_1(\vtheta, \vJ, t) = \sum_{\vn} \hfn(\vJ, t) \e^{i \vn \cdot \vtheta},
  \label{eq:Fourier_expansion_theta}
\end{align}
where $\vn = (\nr,\npsi,\nphi)$ is a vector of integers. $\hPhin(\vJ,t)$ is given in Appendix \ref{sec:app_Psi}. The torque exerted on a single orbit is then
\begin{align}
  \frac{\drm \Lz}{\drm t} &= - \frac{\pd H}{\pd \thetaphi} = - i \sum_{\vn} \nphi \hPhin(\vJ, t) \e^{i \vn\cdot\vtheta},
  \label{eq:torque_single}
\end{align}
and the total torque integrated over the full distribution is
\begin{align}
  &\torq = \int \drm^3 \vJ \drm^3 \vtheta f(\vtheta, \vJ, t) \frac{\drm \Lz}{\drm t}
  \label{eq:torque_fastlimit_1} \\
  &= \!\int\!\drm^3 \vJ \drm^3 \vtheta \left[f_0(\vJ) + \sum_\vn \hfn(\vJ, t) \e^{i \vn\cdot\vtheta}\right] \left[ - i \sum_{\vn'} \nphi' \hPhi_{\vn'}(\vJ, t) \e^{i \vn'\cdot\vtheta}\right]. \nonumber
\end{align}
The first term vanishes upon integration over $\vtheta$ and the second term is non-zero only when $\vn' = - \vn$, so
\begin{align}
  \torq = i \left(2 \pi \right)^3 \sum_\vn \nphi \int \drm^3 \vJ \hfn(\vJ, t) \hPhin^{*}(\vJ, t),
  \label{eq:torque_fastlimit_2}
\end{align}
where we used $\hPhi_{-\vn} = \hPhin^{*}$ as required by ${\rm Im}[\Phi_1]=0$. We see from the above equation that the torque is second order in the perturbation. To proceed, we need to develop the response $\hfn(\vJ, t)$ as a function of the potential perturbation $\hPhin(\vJ, t)$. This is described by the collisionless Boltzmann equation (CBE):
\begin{align}
  \frac{\drm f}{\drm t} = \frac{\pd f}{\pd t} + \frac{\pd f}{\pd \vtheta} \cdot \frac{\pd H}{\pd \vJ} - \frac{\pd f}{\pd \vJ} \cdot \frac{\pd H}{\pd \vtheta} = 0.
  \label{eq:CBE}
\end{align}
Using the Fourier decomposition (\ref{eq:Fourier_expansion_theta}), we have, for each $\vn$,
\begin{align}
  \frac{\pd \hfn}{\pd t} &+ i \vn \cdot \vOmega \hfn - i \vn \cdot \frac{\pd f_0}{\pd \vJ}\hPhin \nonumber \\
  &+ \sum_{\vn'}\left[ i(\vn-\vn') \cdot \frac{\pd \hPhi_{\vn'}}{\pd \vJ} \hf_{\vn - \vn'} - i \vn' \cdot \frac{\pd \hf_{\vn - \vn'}}{\pd \vJ} \hPhi_{\vn'} \right] = 0.
  \label{eq:CBE_fn}
\end{align}
The standard prescription to solve this for $\hfn$ is to ignore the non-linear terms inside the square bracket assuming that both $|f_1/f_0|$ and $|\Phi_1/H_0|$ are sufficiently small. This yields the linearized CBE, which has the following solution for an external perturbation imposed after $t=0$:
\begin{align}
  \hfn(\vJ, t) = i \vn \cdot \frac{\pd f_0}{\pd \vJ} \int_0^t \drm t' \e^{- i \vn \cdot \vOmega (t - t')} \hPhin(\vJ, t').
  \label{eq:fn_linearizedCBE}
\end{align}
Substituting \eqref{eq:fn_linearizedCBE} to (\ref{eq:torque_fastlimit_2}) yields the general torque formula
\begin{align}
  \torq
  =& - \left(2 \pi \right)^3 \sum_\vn \nphi \int \drm^3 \vJ \vn \cdot \frac{\pd f_0}{\pd \vJ} \nonumber \\
  &\times \left[ \int_0^t \drm t' \e^{- i \vn \cdot \vOmega (t - t')} \hPhin(\vJ, t') \right] \hPhin^*(\vJ, t).
  \label{eq:torque_fastlimit_3}
\end{align}
For a perturbation with constant amplitude and pattern speed $\Omegap$, the time dependence separates as $\hPhin(\vJ, t) = \hPhin(\vJ) \e^{-i \nphi \Omegap t}$, and we have \citepalias{Weinberg2004Timedependent}
%
\begin{align}
  \torq = - \left(2 \pi \right)^3  \! \sum_\vn \nphi \! \int \! \drm^3 \vJ \vn \cdot \frac{\pd f_0}{\pd \vJ} |\hPhin|^2 \frac{\sin\left[\left(\vn \cdot \vOmega - \nphi \Omegap\right)t\right]}{\vn \cdot \vOmega - \nphi \Omegap},
  \label{eq:torque_Weinberg}
\end{align}
which, in the time-asymptotic limit $t \rightarrow \infty$, reduces to the LBK formula (\citetalias{Tremaine1984Dynamical}; \citetalias{lynden1972generating})
\begin{align}
  \torq
  =& - \left(2 \pi \right)^3 \sum_\vn \nphi \!\! \int \drm^3 \vJ \vn \cdot \frac{\pd f_0}{\pd \vJ}
  |\hPhin(\vJ)|^2 \pi\delta\left(\vn \cdot \vOmega - \nphi \Omegap\right)\!.
  \label{eq:torque_LBK}
\end{align}
\citetalias{Weinberg2004Timedependent} and more recently \cite{Banik2021SelfConsistent} emphasized the problem of taking the time-asymptotic limit as the age of the galaxy (and yet less that of the perturbation) may not be longer enough than the relevant dynamical time. 

It is obvious from equations (\ref{eq:torque_Weinberg}) and (\ref{eq:torque_LBK}) that the torque is dominated by contributions from phase space near resonances, i.e. $\vn \cdot \vOmega - \nphi \Omegap \approx 0$. However, the vicinity of resonances is precisely where the linearized CBE becomes invalid: if resonances do not move, the linear response (\ref{eq:fn_linearizedCBE}) grows indefinitely at resonances (Appendix \ref{sec:app_linear_response}), and this will soon violate the assumption that the non-linear terms of the CBE (\ref{eq:CBE_fn}) are negligible, which is the basis of this formalism. A correct physical description must respect that the orbits near resonances become trapped, a non-linear secular behaviour which cannot be described by linear theory.

\begin{figure}
  \begin{center}
    \includegraphics[width=8.5cm]{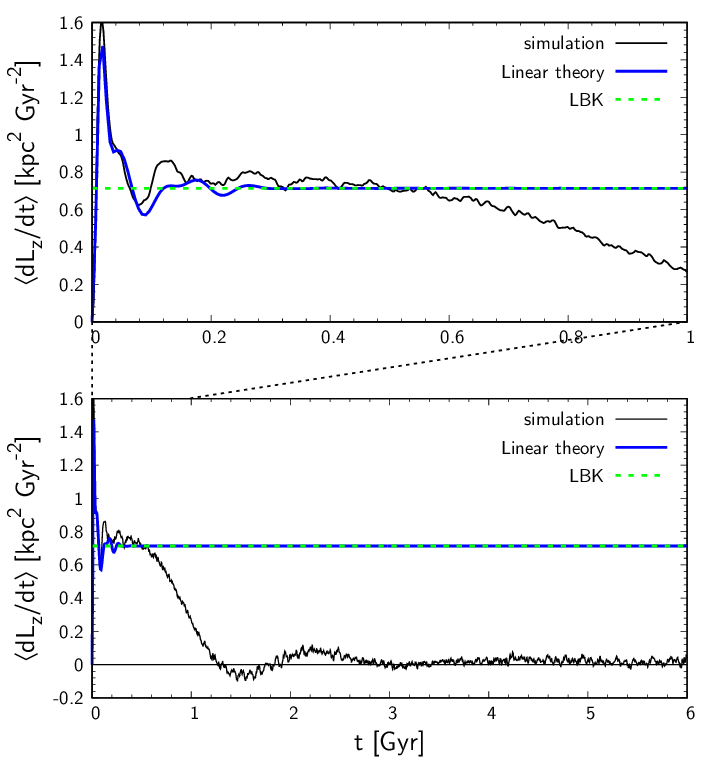}
    \vspace{-1mm}
    \caption{Total torque on the halo (divided by the total halo mass) perturbed by a bar with constant pattern speed and amplitude switched on at $t=0$. Linear theory (blue, equation \ref{eq:torque_Weinberg}) agrees with the torque of the test-particle simulation (black) for a few dynamical period (top panel) but fails to predict the long-term evolution (bottom panel). The LBK formula (green, equation \ref{eq:torque_LBK}) gives the time-asymptotic value of linear theory.}
    \label{fig:torque_W04}
  \end{center}
\end{figure}

To illustrate the problem, we show in Fig.~\ref{fig:torque_W04} the torque exerted on the halo by a constantly rotating bar. We compare the torque measured from a test particle simulation (black) with linear theory (equation~\ref{eq:torque_Weinberg}, blue) and the LBK formula (equation~\ref{eq:torque_LBK}, dashed green). For the two approximations, we restrict the summation to indices with $\nr \in [-10,10]$, $\npsi \in -2,0,2$, and $\nphi = 2$. Since both the halo density and the perturbing potential decrease towards large radii, the contributions are largely limited to the main resonances with small $r$, the strongest being the corotation resonance $\vn = (0,2,2)$. The top panel shows the evolution for the first $1 \Gyr$ which we see a fair agreement between the simulation (black) and the time-dependent linear theory (blue) up to $\sim 500 \Myr$. However, the linear theory qualitatively fails to predict the subsequent long-term evolution shown in the bottom panel: while the torque of linear theory rapidly converges to a non-zero constant value predicted by the LBK formula, the numerical torque fluctuates over a much longer timescale, gradually settling down to zero\footnote{We note that this limit depends on the time-dependence of the perturbation. Here the approach to a steady state (zero torque) is the result of the assumed constant bar pattern speed and amplitude.}. This relaxation is driven by phase mixing (a collisionless relaxation), as we describe in the next section.

In this experiment, we have kept the bar's pattern speed constant. If the bar decelerates sufficiently fast that the resonances pass over the orbits before they can respond non-linearly, resonant trapping will not occur and linear theory will properly predict the torque for an extended period of time as demonstrated in \citetalias{Weinberg2004Timedependent}. For this reason, the linear formalism is said to be valid in the \textit{fast limit} \citepalias[][]{Tremaine1984Dynamical}. The validity is characterized by the speed parameter\footnote{In the limit of epicycle approximation, the speed parameter reduces to $s \sim \eta/A$ where $\eta\equiv-\dOmegap/\Omegap^2$ is the bar's dimensionless slowing rate and $A$ the dimensionless bar strength \citep[][equation 27]{Chiba2020ResonanceSweeping}.}\citepalias{Tremaine1984Dynamical}:
\begin{align}
  s \equiv \left\vert \frac{\nphi \dOmegap}{\omega_0^2} \right\vert,
  \label{eq:speed_parameter}
\end{align}
where $\dOmegap$ measures the long-term evolution of the pattern speed \citep[i.e. not the short-term fluctuations as discussed in][]{Wu2016TimeDependent} and $\omega_0$ is the libration frequency of trapped orbits at the core of the resonance (equation \ref{eq:w0_J0_eps}). In the fast limit $s \gg 1$, orbits cannot stay trapped in resonance because the resonant potential no longer features a local minimum \citep[\citetalias{Tremaine1984Dynamical},][see also Appendix \ref{sec:app_orbit_subject_to_slowing_bar}]{Chiba2020ResonanceSweeping}.

If the perturbation vanishes before any non-linear effect develops, the formula will be valid regardless of $s$. This could be the case for e.g. transient spiral arms. Hence, the traditional torque formulas (\ref{eq:torque_fastlimit_3})-(\ref{eq:torque_LBK}) are appropriate to study the evolution of perturbations that either decelerate rapidly or decay rapidly. Since neither applies to long-lived galactic bars, we must fully model the non-linear response of the halo, i.e. trapping at resonances.

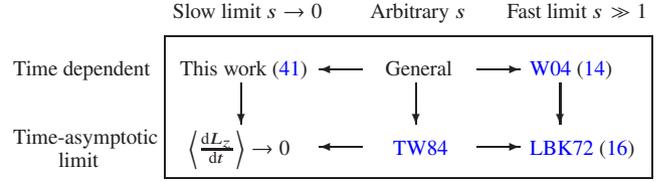
\begin{figure}
  \setlength{\unitlength}{1mm}
  \centering
  \begin{picture}(100,24)
    \put(0,13) {Time dependent}
    \put(0,4) {Time-asymptotic}
    \put(6,1) {limit}
    \put(21,20.5) {Slow limit $s \rightarrow 0$}
    \put(47,20.5) {Arbitrary $s$}
    \put(65,20.5) {Fast limit $s \gg 1$}
    \put(22,13) {This work (\ref{eq:torque_slow_regime})}
    \put(49,13) {General}
    \put(68,13) {\citetalias{Weinberg2004Timedependent} (\ref{eq:torque_fastlimit_3})}
    \put(23,2.5) {$\torq \rightarrow 0$}
    \put(50,2.5) {\citetalias{Tremaine1984Dynamical}}
    \put(68,2.5) {\citetalias{lynden1972generating} (\ref{eq:torque_LBK})}
    \put(30,12){\vector(0,-1){5.5}}
    \put(53,12){\vector(0,-1){5.5}}
    \put(72,12){\vector(0,-1){5.5}}
    \put(46,13.7){\vector(-1,0){5.7}}
    \put(46,3.5){\vector(-1,0){5.7}}
    \put(61,13.7){\vector(1,0){5.7}}
    \put(61,3.5){\vector(1,0){5.7}}
    \put(20,-0.5){\framebox(64,18.5)}
  \end{picture}
  \vspace{-1mm}
  \caption{Domain of applicability. The numbers in brackets numerate the equations within our paper. \citetalias{Tremaine1984Dynamical} formulated the torque for arbitrary $s$ in the time-asymptotic limit, and demonstrated the recovery of the \citetalias{lynden1972generating} formula (\ref{eq:torque_LBK}) in the fast limit ($s \gg 1$). The time-dependent formula (\ref{eq:torque_fastlimit_3}) derived by \citetalias{Weinberg2004Timedependent} is valid in the fast limit and reduces to LBK in the time-asymptotic limit. We formulate the time-dependent torque in the slow limit ($s \rightarrow 0$) which, in the time-asymptotic limit, predicts zero torque in agreement with \citetalias{Tremaine1984Dynamical}.}
  \label{fig:domain_applicability}
\end{figure}

\section{Torque in the slow limit}
\label{sec:torque_slow_regime}

\subsection{Approach by Tremaine \& Weinberg (1984)}
\label{sec:TW84}

The torque in the slow regime has been discussed in \citetalias{Tremaine1984Dynamical}, who derived a general formula of dynamical friction valid for arbitrary speed parameter $s$ by Taylor expansion around each sweeping resonance (equation 86 in \citetalias{Tremaine1984Dynamical}). By retaining the first two terms of the Taylor series, they demonstrate that the \citetalias{lynden1972generating} formula (\ref{eq:torque_LBK}) is recovered in the fast limit $(s \gg 1)$. In the slow non-linear regime $(s < 1)$, by retaining the leading order term, \citetalias{Tremaine1984Dynamical} approximates the total dynamical friction by the torque on resonant orbits averaged over their initial phase and integrated from $t=-\infty$ to $t=\infty$. The resulting equation\footnote{A summarized formula is given in equation (39) of \cite{weinberg1985evolution}.} describes the net change in angular momentum of orbits jumped over a moving resonance. This slow formula, however, does not describe the temporal fluctuations in the torque caused by the transient responses to the passage of a resonance. As we show in section \ref{sec:slowdown_bar}, the sweeping resonance leaves behind a striated perturbation in phase space that requires several orbital periods to phase mix and reach a steady state. The slow formula of \citetalias{Tremaine1984Dynamical} is thus equivalent in essence to the prediction in the time-asymptotic limit, i.e. the net torque after any phase imbalance has smoothed away by phase mixing. In the slow limit ($s \rightarrow 0$ and thus $\dOmegap \rightarrow 0$), the \citetalias{Tremaine1984Dynamical} formula predicts zero torque (since their torque depends linearly on $\dOmegap$) which is consistent with the limiting behaviour of the damped oscillation shown in Fig.~\ref{fig:torque_W04}. We summarize the relations among the existing theories in Fig.~\ref{fig:domain_applicability}. The formulae are classified based on slow vs. fast limit and time-dependent vs. time-asymptotic limit. In this map, our equation (\ref{eq:torque_slow_regime}) derived in the following fills the top-left corner: a time-dependent formula valid in the slow limit ($s \rightarrow 0$). By our notation $s \rightarrow 0$, we imply the strict limit, where the bar rotates constantly, and not a bar slowing in the limit of small $s$. The time-dependent dynamical friction in the general slow regime $(0 < s < 1)$ will be discussed numerically in section \ref{sec:slowdown_bar}.

\subsection{Angle-action variables near resonances}
\label{sec:angle_action_nonaxisymmetric_potential}

The alternative to Taylor expansion or to solving the non-linear terms in the CBE (\ref{eq:CBE_fn}) is simply to change the coordinates. The majority of orbits near each resonance $\vN=(\Nr,\Npsi,\Nphi)$ behave quasi-periodically regardless of trapping and thus there exists a new set of angle-action coordinates for them \citep[e.g. \citetalias{Tremaine1984Dynamical};][]{Kaasalainen1994hamiltonian,Sridhar1996Adiabatic,Binney2016Managing,binney2017orbital,monari2017distribution}. In these resonant angle-action coordinates, the time evolution of the distribution function can be solved trivially using the non-linearized CBE (section \ref{sec:evolution_of_DF}). In this section, we first lay out the known coordinate transformations, and from there develop our equations for the torque on both trapped and untrapped orbits.

We start with a canonical transformation to the so-called slow-fast angle-action variables $(\vtheta',\vJ')=(\thetafo,\thetaft,\thetas,\Jfo,\Jft,\Js)$ \citepalias[e.g.][]{Tremaine1984Dynamical}:
\begin{align}
  &\thetafo = \thetar, \hspace{14.5mm} \thetaft = \thetapsi, \hspace{13mm} \thetas  = \vN \cdot \vtheta - \Nphi \Omegap t,
  \label{eq:slowfast_angle} \\
  &\Jfo = \Jr - \frac{\Nr}{\Nphi}\Lz, \hspace{3mm} \Jft = L - \frac{\Npsi}{\Nphi}\Lz, \hspace{3mm} \Js  = \frac{\Lz}{\Nphi},
\label{eq:slowfast_action}
\end{align}
using the generating function
\begin{align}
  W(\vtheta,\vJ',t) = \thetar \Jfo + \thetapsi \Jft + \left(\vN \cdot \vtheta - \Nphi \Omegap t\right) \Js.
\label{eq:generating_function}
\end{align}
The Hamiltonian (\ref{eq:H0_H1_f0_f1}) transforms to
\begin{align}
  H(\vtheta', \vJ')
  &= H_0(\vJ') - \Nphi \Omegap \Js + \sum_{\vk} \hPsik(\vJ') ~\e^{i \vk \cdot \vtheta'},
  \label{eq:Hamiltonian_FourierExpand}
\end{align}
where we have expanded the perturbation $H_1$ into a Fourier series (Appendix \ref{sec:app_Psi}) with indices $\vk=(\kfo,\kft,\ks)$. The purpose of transforming to the slow-fast variables is to separate the motion into slow and fast components: near the resonance, the slow angle $\thetas$ evolves much slower than the fast angles $\vthetaf = (\thetafo,\thetaft)$. This allows one to average the Hamiltonian over $\vthetaf$ while holding $\thetas$ fixed which removes all perturbation terms in (\ref{eq:Hamiltonian_FourierExpand}) with $\vkf = (\kfo,\kft) \neq {\bm 0}$. Since only terms with $\ks = \pm 1$ are non-zero for resonances with $\Nphi = 2$ (Appendix \ref{sec:app_Psi}), we may write
\begin{align}
  \bH(\thetas,\vJ')
  &= H_0(\vJ') - \Nphi \Omegap \Js + \Psi(\vJ') \cos \left(\thetas - \thetasres\right),
  \label{eq:Hamiltonian}
\end{align}
where $\Psi \equiv 2|\hPsi_{(0,0,1)}|$. In this averaged system, the two fast actions $\vJf = (\Jfo,\Jft)$ are conserved, i.e. $\dot{\bm J}_{\rm f} = - \frac{\partial \bH}{\partial \vthetaf} = {\bm 0}$. Thus this Hamiltonian is integrable: it has three isolating integrals, the two fast actions and the Hamiltonian itself, meaning that angle-action coordinates for $\bH$ exist.

\begin{figure}
  \begin{center}
    \includegraphics[width=8.5cm]{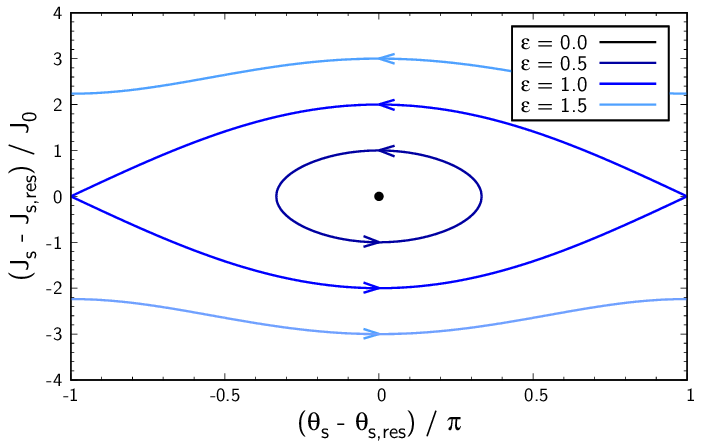}
    \vspace{-1mm}
    \caption{Phase flow for the approximate $\bH$ (\ref{eq:Hamiltonian_Taylor_expand}). Level curves of $\bH$ parametrized by $\varepsilon$ are plotted. For $0 \leq \varepsilon < 1$, the motion of orbits in $\thetas$ is bounded (libration) whereas for $\varepsilon > 1$, $\thetas$ is unbounded (circulation).}
    \label{fig:slowAA_schematic}
  \end{center}
\end{figure}

A convenient analytical transformation to the angle-action coordinates of $\bH$ is available if we Taylor expand $\bH$ around the resonance $\Jsres$ up to second order and ignore terms smaller than $|\Psi/H_0|$:
%
\begin{align}
  \bH(\thetas,\vJ')
  \simeq& \frac{1}{2} G \left(\vJf,\Jsres\right) \left(\Js - \Jsres\right)^2 \nonumber \\
  &+ \Psi \left(\vJf,\Jsres\right) \cos \left(\thetas - \thetasres\right),
  \label{eq:Hamiltonian_Taylor_expand}
\end{align}
where 
\begin{align}
  G\left(\vJ'\right) \equiv \frac{\partial^2 H_0}{\partial \Js^2} = \frac{\partial \Omegas}{\partial \Js}.
  \label{eq:G}
\end{align}
The first order derivative of $H_0(\vJ') - \Nphi \Omegap \Js$ vanishes at the resonance and we have dropped the constant terms. The approximated Hamiltonian takes the form of a classical pendulum Hamiltonian apart from a different sign. In galactic dynamics, $G$ (equation \ref{eq:G}) is in many cases negative \citepalias[][Figure 5]{Tremaine1984Dynamical}, so in what follows we will assume $G < 0$ although cases with $G > 0$ can be treated similarly by adjusting $\thetasres$ as appropriate. For subsequent use, we define
\begin{align}
  \omega_0^2 \equiv -G \Psi, ~~~ J_0^2 \equiv - \frac{\Psi}{G}, ~~~ \varepsilon \equiv \sqrt{\frac{1}{2}\left(1 - \frac{\bH}{\Psi}\right)}.
  \label{eq:w0_J0_eps}
\end{align}
Depending on the value of $\bH$, parametrized by the dimensionless energy of the pendulum $\varepsilon$, there are two types of qualitatively different motions as depicted in Fig.~\ref{fig:slowAA_schematic}: orbits librate when $0 \leq \varepsilon < 1$ (trapped) and circulate above or below the resonance when $\varepsilon > 1$ (untrapped). $\varepsilon=1$ defines the separatrix.

In the following, we summarize the canonical transformation from the slow angle-action variables $(\thetas,\Js)$ to the angle-action coordinates of libration/circulation \citep[][]{Brizard2013Jacobi}, and formulate the secular torque on trapped/untrapped orbits.

\subsubsection{Trapped orbits ($0 \leq \varepsilon < 1$)}
\label{sec:AA_lib}

The angle-action coordinates of libration are
\begin{align}
  &\thetal = \theta_{\ell 0} + \Omegal t, ~~ \Omegal = \frac{\pi}{2 K(\varepsilon)} \omega_0,
  \label{eq:thetal_omegal} \\
  &\Jl = \frac{1}{2\pi} \oint \drm \thetas \Js = \frac{8 J_0}{\pi} \left[ E(\varepsilon) - \left(1 - \varepsilon^2 \right) K(\varepsilon) \right],
  \label{eq:Jl}
\end{align}
where $K(\varepsilon)$ and $E(\varepsilon)$ are the complete elliptic integral of the first and second kinds (Appendix \ref{sec:app_elliptic_functions}). $\omega_0$ and $J_0$ are defined in equation (\ref{eq:w0_J0_eps}). The inverse transformation is 
\begin{align}
  &\thetas - \thetasres = 2 \arcsin \left[ \varepsilon {\rm sn} \left(\left.\frac{2 K(\varepsilon)}{\pi} \thetal \right\vert \varepsilon \right)\right], \\
  &\Js - \Jsres = 2 J_0 \varepsilon {\rm cn} \left(\left.\frac{2 K(\varepsilon)}{\pi} \thetal \right\vert \varepsilon \right),
  \label{eq:libAA_to_slowAA}
\end{align}
where sn, cn and dn are the Jacobi elliptic functions (Appendix \ref{sec:app_elliptic_functions}). The torque exerted on trapped orbits is then
\begin{align}
  \frac{\drm \Lz}{\drm t}
  &= \Nphi \Omegal \frac{\drm \Js}{\drm \thetal}
  = 2 \Nphi \varepsilon \Psi {\rm sn}\!\left(\!\left.\frac{2 K(\varepsilon)}{\pi} \thetal \right\vert \varepsilon \right) {\rm dn}\!\left(\!\left.\frac{2 K(\varepsilon)}{\pi} \thetal \right\vert \varepsilon \right),
  \label{eq:torque_lib}
\end{align}
which becomes harmonic near the centre of the resonance:
\begin{align}
  \frac{\drm \Lz}{\drm t} \simeq 2 \Nphi \varepsilon \Psi \sin \thetal ~~~~ {\rm for} ~~~~ \varepsilon \ll 1.
  \label{eq:torque_lib_limit}
\end{align}

\subsubsection{Non-trapped orbits ($\varepsilon > 1$)}
\label{sec:AA_rot}

The angle-action coordinates of circulation are
\begin{align}
  &\thetac = \theta_{\rm c 0} + \Omegac t,~~ \Omegac = \frac{\pi \varepsilon}{K(\varepsilon^{-1})} \omega_0,
  \label{eq:thetac_omegac} \\
  &\Jc = \frac{1}{2\pi} \oint \drm \thetas \Js = \Jsres \pm \frac{4 J_0 \varepsilon}{\pi} E(\varepsilon^{-1}),
  \label{eq:Jc}
\end{align}
where the $\pm$ signs correspond to the upper and lower circulating regimes, respectively. Note that the subscript `c' denotes `circulating' (untrapped) orbits and not `circular' orbits. The inverse transformation is 
\begin{align}
  &\thetas - \thetasres = \pm 2 \arcsin \left[ {\rm sn} \left(\left.\frac{K(\varepsilon^{-1})}{\pi} \thetac \right\vert \varepsilon^{-1} \right)\right], \\
  &\Js - \Jsres = \pm 2 J_0 \varepsilon {\rm dn} \left(\left.\frac{K(\varepsilon^{-1})}{\pi} \thetac \right\vert \varepsilon^{-1} \right).
  \label{eq:rotAA_to_slowAA}
\end{align}
In the limit $\varepsilon \rightarrow \infty$, the slow action is conserved and the slow angle $\thetas$ asymptotically approaches the angle of circulation $\thetac$. The torque exerted on untrapped orbits is
\begin{align}
  \frac{\drm \Lz}{\drm t}
  &= \pm 2 \Nphi \Psi {\rm sn} \left(\left.\frac{K(\varepsilon^{-1})}{\pi} \thetac \right\vert \varepsilon^{-1} \right) {\rm cn} \left(\left.\frac{K(\varepsilon^{-1})}{\pi} \thetac \right\vert \varepsilon^{-1} \right).
  \label{eq:torque_rot}
\end{align}
Far from the resonance, the torque again becomes harmonic:
\begin{align}
  \frac{\drm \Lz}{\drm t} \simeq \pm \Nphi \Psi \sin \thetac ~~~~ {\rm for} ~~~~ \varepsilon \gg 1.
  \label{eq:torque_rot_limit}
\end{align}

\begin{figure}
  \begin{center}
    \includegraphics[width=8.5cm]{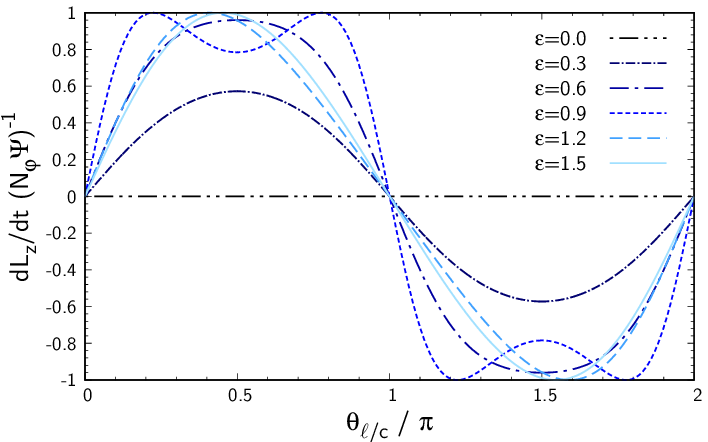}
    \vspace{-1mm}
    \caption{Torque applied to trapped $(\varepsilon < 1)$ and untrapped $(\varepsilon > 1)$ orbits as a function of the libration/circulation angle $\thetalc$. The behaviour deviates from a sinusoidal wave near the separatrix $\varepsilon = 1$.}
    \label{fig:dLzdt_thetap}
  \end{center}
\end{figure}

The behaviour of the torque is plotted in Fig.~\ref{fig:dLzdt_thetap}. As seen in (\ref{eq:torque_lib_limit}) and (\ref{eq:torque_rot_limit}), the torque fluctuates sinusoidally near the core $(\varepsilon \ll 1)$ of the resonance, as well as far from it $(\varepsilon \gg 1)$. The torque on trapped orbits diminishes towards $\varepsilon = 0$ where it vanishes. Near the separatrix $(\varepsilon \approx 1)$, the torque forms a dip at $\thetal = \pi/2, 3\pi/2$ since orbits stagnate near the unstable singular points $(\thetas - \thetasres, \Js - \Jsres) = (\pm \pi, 0)$, which, for the corotation resonance, correspond to the unstable Lagrange points $L_{1,2}$ on the bar's major axis. As the orbit approaches the separatrix, the dip deepens and eventually the torque becomes zero over the entire phase except at $\thetal = 0$ and $\pi$. This situation is analogous to a real pendulum in transition from oscillation to circulation that spends most of its time/phase stagnating at the top of the pivot with hardly any change in momentum.

\subsection{Evolution of the distribution function}
\label{sec:evolution_of_DF}

The averaged Hamiltonian $\bH$ (\ref{eq:Hamiltonian_Taylor_expand}) now only depends on the new actions $\vJ'' = (\Jfo,\Jft,\Jlc)$ that are conjugate to the new angles $\vtheta'' = (\thetafo,\thetaft,\thetalc)$ where the subscript `$\ell/{\rm c}$' indicates libration or circulation depending on $\varepsilon$. The collisionless Boltzmann equation (\ref{eq:CBE}) for the averaged system is then 
\begin{align}
  \frac{\pd f}{\pd t} + \vOmega'' \cdot \frac{\pd f}{\pd \vtheta''} = 0, ~~{\rm where}~~~ \vOmega''(\vJ'') \equiv \frac{\pd \bH}{\pd \vJ''}.
  \label{eq:CBE_integrable}
\end{align}
This equation has the trivial solution
\begin{align}
  f(\vtheta'',\vJ'',t) = f(\vtheta'' - \vOmega'' t,\vJ'',0).
  \label{eq:f2}
\end{align}
One can thus readily obtain the phase space density at any time $t = \tau$ from the original distribution at $t = 0$ by simply winding the angles backward by $\vOmega'' \tau$. Since the initial unperturbed distribution of the halo (\ref{eq:app_Hernquist_DF}) is uniform in the fast angle $(\thetafo,\thetaft)$ but \textit{not} in the libration/circulation angle $\thetalc$, the system is out of equilibrium after the emergence of the bar and starts phase mixing in $\thetalc$.

\begin{figure*}
  \begin{center}
    \setlength\columnsep{-40pt}
    \begin{multicols}{2}
      \includegraphics[width=8.0cm]{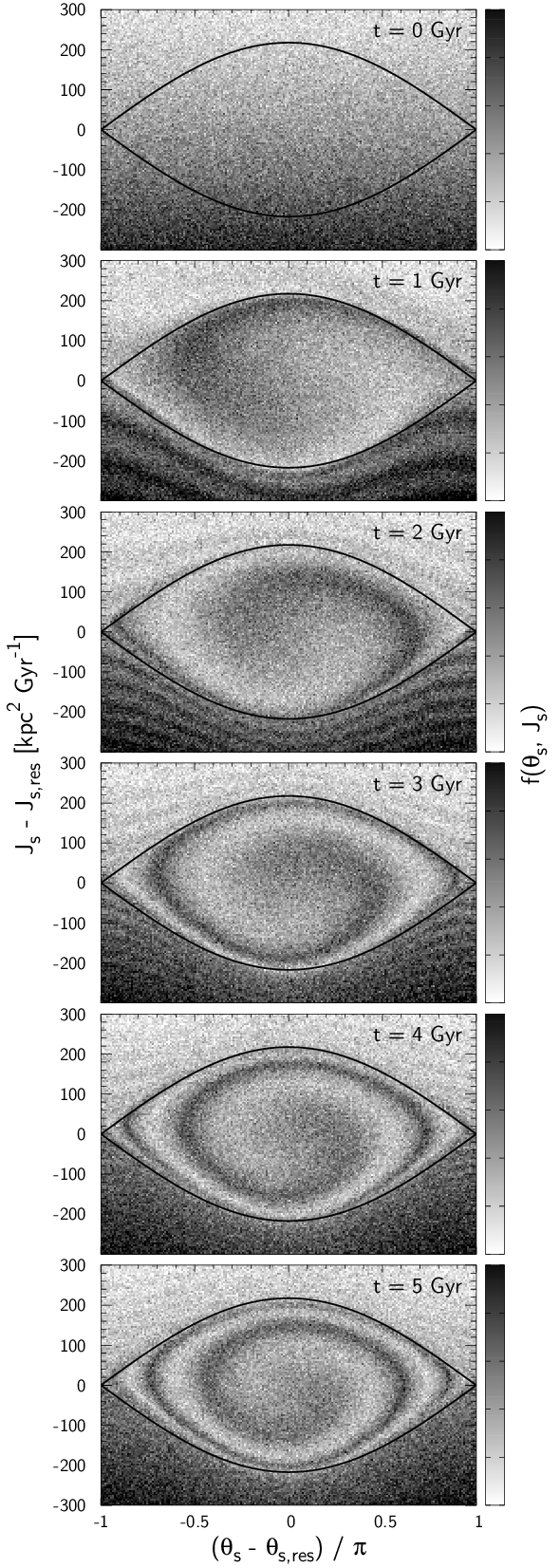}
      \newpage
      \includegraphics[width=8.0cm]{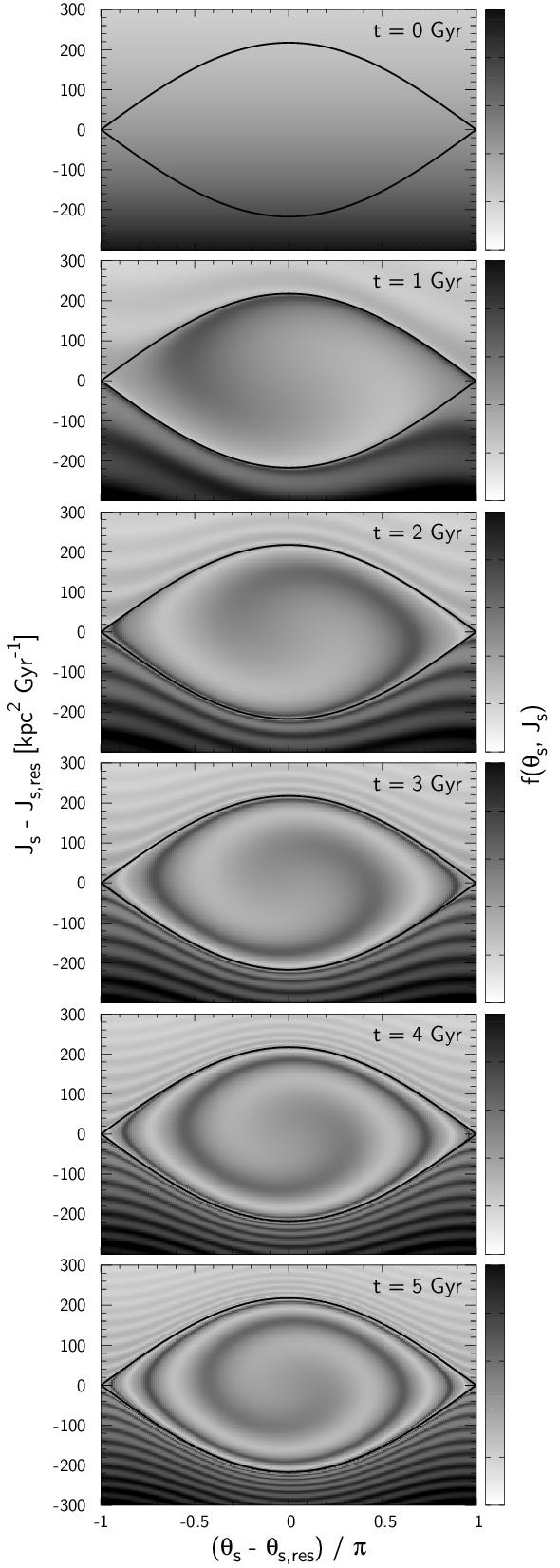}
    \end{multicols}
    \vspace{-4mm}
    \caption{Time evolution of the phase-space density $f(\thetas,\Js)$ near the corotation resonance $\vN=(0,2,2)$ at $(\Jfo,\Jft) = (10,0) \kpckpcGyr$ plotted every $1 \Gyr$. Left column: test particle simulation. Right column: analytical (equation \ref{eq:f2}). Black curves mark the separatrix.}
    \label{fig:phase_flow_slowAA}
  \end{center}
\end{figure*}

Figure~\ref{fig:phase_flow_slowAA} shows the evolution of the phase-space density near the corotation resonance (CR) at $(\Jfo,\Jft) = (10,0) \kpckpcGyr$ plotted every $1 \Gyr$ from top to bottom. The bar is instantaneously switched on at $t=0$. The left-hand column shows the distribution of the test-particle simulation which is well reproduced by the analytical distribution shown on the right. Inside the separatrix (black curves), the density, initially uniform in the slow angle, phase mixes in the libration angle. Since the libration period increases towards the separatrix, the phase space differentially rotates, resulting in a phase-space spiral. Similarly, the phase space outside the separatrix winds into stripes since the circulation period is also maximal (infinite) at the separatrix and drops away from there.

The general mechanics of this phase spiral is rather similar to the vertical phase spiral found in \textit{Gaia} \citep{Antoja2018Nature}. In the case of the Gaia phase-spiral, a non-uniform distribution in the vertical angle is created, e.g., by the impact of a dwarf galaxy \citep{Binney2018GaiaPhaseSpiral,BlandHawthorn2021GalacticSeismology}. In our case, the non-uniformity in the libration and circulation angles is inherited from the phase-space configuration prior to bar formation. We also remark that the phase mixing of trapped orbits and its resulting decay in dynamical friction is equivalent in mechanism to the non-linear Landau damping observed in plasma systems \citep{ONeil1965NonlinearLandauDamping,Malmberg1967Collisionless}.

\begin{figure}
  \begin{center}
    \includegraphics[width=7.9cm]{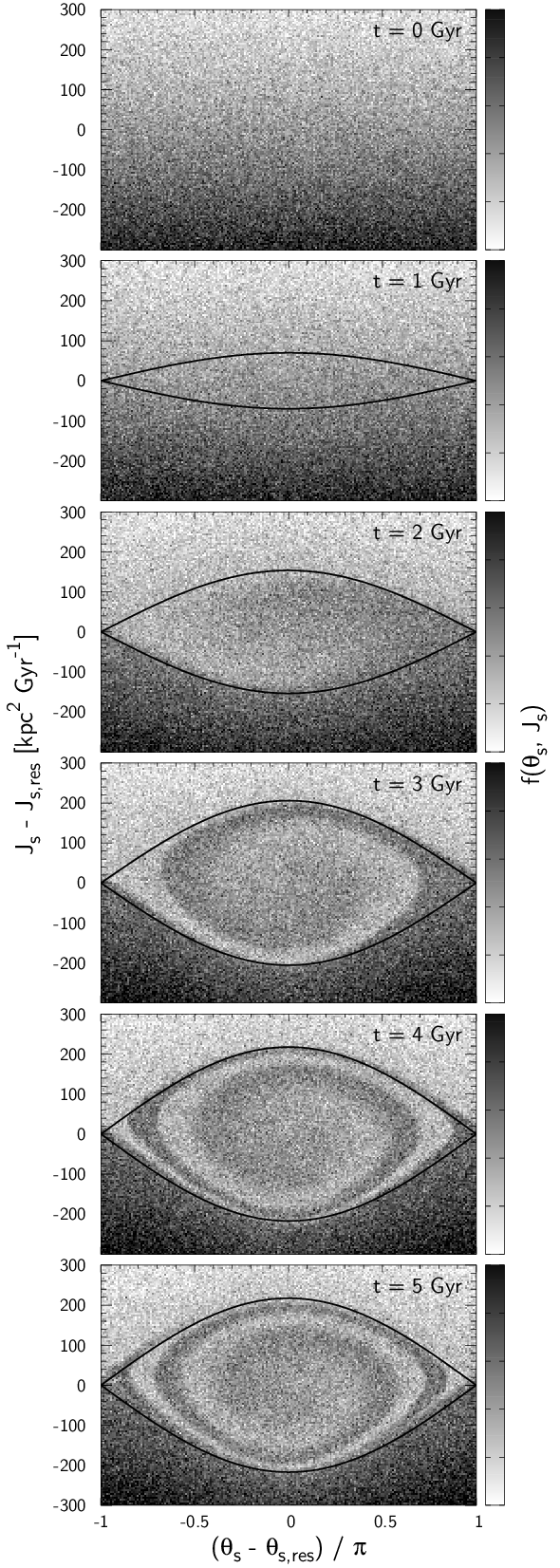}
    \vspace{-2mm}
    \caption{Phase-space density of the halo perturbed by a bar growing slowly in $4 \Gyr$. The distribution outside the resonance is smooth due to the adiabatic emergence of the bar, while that inside the resonance displays a winding spiral since the expanding separatrix simultaneously captures orbits from above (low density) and below (high density).}
    \label{fig:phase_flow_slowAA_grow}
  \end{center}
\end{figure}

The development of a phase-space spiral inside the resonance does not hinge on the instantaneous emergence of the bar. It also happens in a slowly growing bar, as shown in Fig.~\ref{fig:phase_flow_slowAA_grow}, which displays the same phenomenon for a bar that slowly grows over $4 \Gyr$ following a polynomial prescription \citep[][]{dehnen2000effect}. Here, we resort to test-particle simulations only because our current analytical treatment (\ref{eq:f2}) assumes conservation of $\Jlc$, which is violated at the separatrix when the bar grows, leading to resonant capture. Far outside the separatrix, the change is generally adiabatic since the circulation periods are small compared to the timescale on which the potential changes. There, the circulation action $\Jc$ is conserved (see Appendix \ref{sec:app_conservation_Jc}) and so the distribution in the circulation angle is kept uniform as it was at $t=0$, thus showing no stripes. The evolution in the trapped region is, however, qualitatively different. The separatrix simultaneously engulfs the high-density region from below and the low-density region from above, thus rendering the distribution in the libration angle inevitably nonuniform (almost like a rectangular wave distribution). Due to this abrupt transition between maximum and minimum density, the phase-space spiral in a growing bar has a hard boundary compared to the non-growing case (Fig.~\ref{fig:phase_flow_slowAA}) which shows a smoother transition between peaks and troughs.

The conservation of $\Jc$ implies that the untrapped orbits bend around the advancing separatrix: near the midpoint $\thetas - \thetasres = 0$ they bend away from the resonance line, while near $\thetas - \thetasres = \pm \pi$ they bend towards the resonance. In the end, the orbits just above and below the separatrix originate from around $\Js - \Jsres \approx \pm 120 \kpckpcGyr$, respectively, which leads to the discontinuous step in density at the resonance when moving along $\Js$ at $\thetas - \thetasres = \pm \pi$.

\begin{figure}
  \begin{center}
    \includegraphics[width=8.5cm]{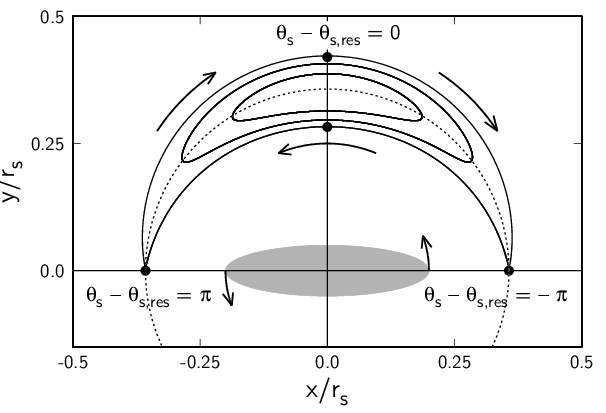}
    \vspace{-1mm}
    \caption{Guiding orbits of particles in the galactic plane ($\beta=0$) trapped in the CR of the bar (gray ellipse). Resonantly trapped orbits at $\thetas - \thetasres > 0 ~ (x < 0)$ are pulled forward by the bar (gain angular momentum), while those at $\thetas - \thetasres < 0 ~ (x > 0)$ are pulled back by the bar (lose angular momentum). The net torque on the bar is determined by the phase distribution of these trapped dark matter. As they slowly phase mix, the net torque attenuates.}
    \label{fig:xy}
  \end{center}
\end{figure}

It is instructive to relate the behaviour in the slow angle-action plane to the Chandrasekhar's classical picture of dynamical friction, i.e. as an over-density lagging behind the moving object (bar). At the CR, the slow angle (\ref{eq:slowfast_angle}) is $\thetas = 2[\thetapsi - (\phib - \thetaphi)]$ where $\thetapsi$ is the azimuthal angle of the guiding orbit in its orbital plane, and $\phib - \thetaphi$ is the azimuthal angle of the bar in the galactic plane measured from the ascending node. The centre of oscillation of the CR is at the bar's minor axis $\thetapsi - (\phib - \thetaphi) = \pi/2$, so $\thetasres = \pi$, and thus the range $\thetas - \thetasres \in [-\pi,\pi]$ corresponds to $\thetapsi - (\phib - \thetaphi) \in [0,\pi]$. Figure~\ref{fig:xy} shows typical in-plane $(\beta = 0)$ orbits in the bar's corotating frame. For $\thetas - \thetasres > 0$, the trapped orbits are lagging behind the bar, so they exert a negative torque on the bar, while for $\thetas-\thetasres < 0$, they are ahead of the bar, exerting a positive torque on the bar. The net torque is then determined by the distribution in $\thetas$: In the beginning, the overdensity at small $z$-angular momentum (or $\Js$) librates towards $\thetas - \thetasres > 0$ thus giving rise to a net negative torque on the bar. Subsequently, this overdensity librates back and forth between $\thetas - \thetasres > 0$ and $\thetas - \thetasres < 0$, providing a systematic oscillation of dynamical friction, which will fade with phase mixing. Eventually, the distribution in $\thetas$ approaches equilibrium and the net torque on a constantly rotating bar converges to zero as we saw in Fig.~\ref{fig:torque_W04}.

If the perturbation is transient (e.g. short-lived spiral arms), once it fades away, the halo will resume phase mixing in the slow angle $\thetas$, which recovered its role as an angle variable. However, since the libration angle $\thetal$ at which the orbits were captured and released are generally different, the final DF near resonances will be flatter than the original distribution. This `scar' in the DF will affect the evolution of the renewed spiral arms \citep{SellwoodCarlberg2014,Sridhar2019Renewal}, and can cause self-gravitational instabilities, exciting further spiral modes \citep{Sellwood2019SpiralInstabilities}.

\subsection{Onset of chaos by overlap of resonances}
\label{sec:overlap_of_resonances}

\begin{figure*}
  \begin{center}
    \includegraphics[width=17.5cm]{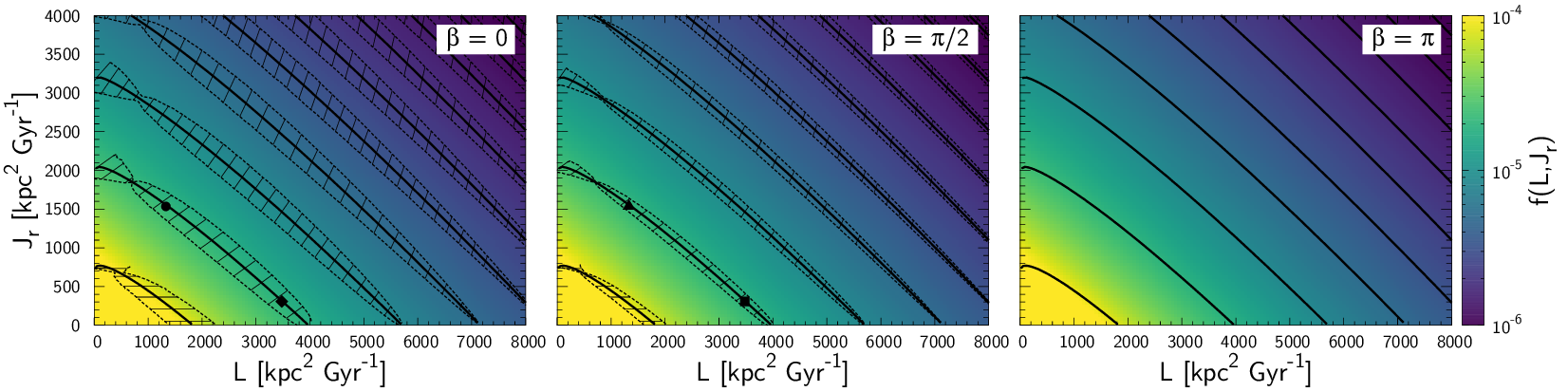}
    (a)~Resonances with $\Npsi = 2, \Nphi=2$. $\Nr$ spans from 0 (left bottom) to 9 (top right).
    \includegraphics[width=17.5cm]{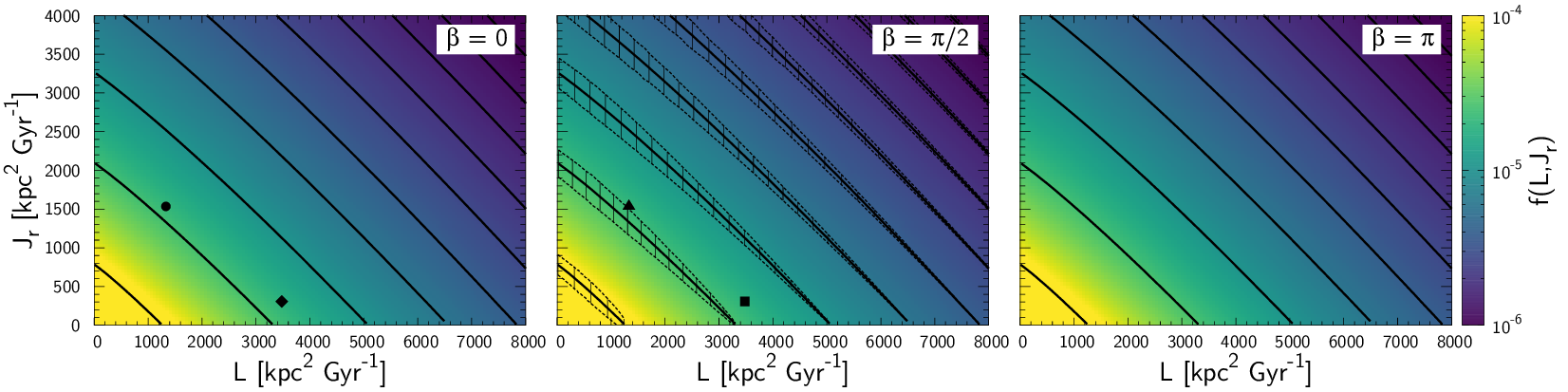}
    (b)~Resonances with $\Npsi = 0, \Nphi=2$. $\Nr$ spans from 1 (left bottom) to 10 (top right).
    \includegraphics[width=17.5cm]{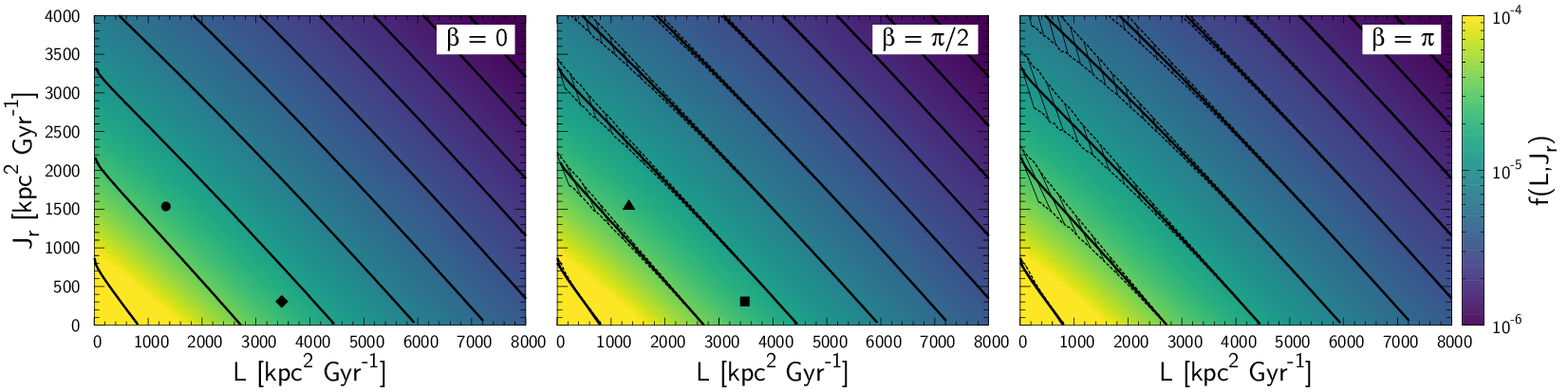}
    (c)~Resonances with $\Npsi = -2, \Nphi=2$. $\Nr$ spans from 2 (left bottom) to 11 (top right).
    \caption{Bar-halo resonances $\vN = (\Nr,\Npsi,\Nphi)$ at three different orbital planes $\beta = \cos^{-1}(\Lz/L) = 0, \pi/2, \pi$ from left to right. The thick black curves mark the location of resonances, the dotted curves mark the maximum extent of the separatrix, and the thin black lines mark the direction of libration. The colours show the phase-space density in log scale. Top panel: Resonances with $\Npsi = 2$. Resonantly trapped volume is largest at $\beta = 0$ and vanishes at $\beta = \pi$. The black symbols, placed on the outer Lindblad resonance $\vN=(1,2,2)$, mark the initial actions of orbits shown in Fig.~\ref{fig:slowAA_overlap}. Middle panel: Resonances with $\Npsi = 0$. Trapped volume is largest at $\beta = \pi/2$ and decays towards $\beta = 0,\pi$. Bottom panel: Resonances with $\Npsi = -2$. Trapped volume increases towards $\beta = \pi$.}
    \label{fig:f_res_J}
  \end{center}
\end{figure*}

By averaging over the fast motions, we have approximated the Hamiltonian near a resonance with an integrable Hamiltonian $\bH$ (\ref{eq:Hamiltonian_Taylor_expand}). We now ask whether the system remains integrable when the neglected terms of the full Hamiltonian $\delta H = H - \bH$ are added back in. To consider the effect of $\delta H$, we Fourier expand it in the new angles
\begin{align}
  \delta H(\vtheta'',\vJ'') = \sum_{\vm} \hPsim(\vJ'') \e^{i \vm \cdot \vtheta''}.
  \label{eq:torque_slow_regime}
\end{align}
The Kolmogorov-Arnold-Moser (KAM) theorem \citep{Arnold1963KAM} implies that these terms will only slightly perturb the motion and the system will stay integrable so long as they are small and sufficiently far from satisfying a resonance condition $\vm \cdot \vOmega'' = 0$ where $\vOmega'' = (\Omegafo,\Omegaft,\Omegalc)$. However, if $\delta H$ contains a resonant term, it may render the system non-integrable. These \textit{secondary resonances} form new resonant islands (either inside or outside the original separatrix) and can in principle be modelled in much the same way as we have dealt with the original resonance of the unperturbed system $H_0$ \citep[e.g.][]{lichtenberg1992regular,Malhotra1998OrbitalResonances,Wisdom2004SpinOrbit}. However, when the separatrices of the secondary resonances come sufficiently close to the separatrix of the original resonances, orbits may behave stochastically, moving from one domain of resonance to another \citep[][]{chirikov1979universal}. A significant secondary resonance is expected when two resonances of the unperturbed Hamiltonian $H_0$ reside close to each other. The resonances in a bar-halo system indeed partially overlap, and some orbits there turn chaotic.

\begin{figure}
  \begin{center}
    \includegraphics[width=8.5cm]{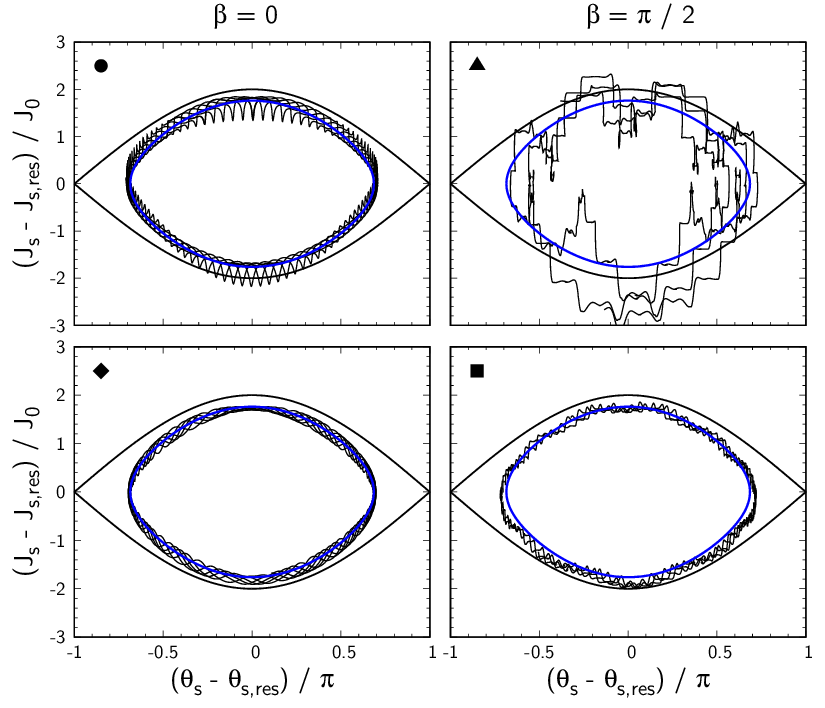}
    \vspace{-1mm}
    \caption{Motion of orbits trapped at the OLR. The symbols on the left top corner represent the position of the initial actions marked in Fig.~\ref{fig:f_res_J}. The black orbits are integrated numerically using the full Hamiltonian and the blue orbits are obtained analytically from the averaged Hamiltonian. The orbit in the top-right panel exhibits chaotic oscillations as it is overlapped by the $\vN=(2,0,2)$ resonance. All orbits have similar values of $\varepsilon \sim 0.9$.}
    \label{fig:slowAA_overlap}
  \end{center}
\end{figure}

\begin{figure}
  \begin{center}
    \includegraphics[width=8.5cm]{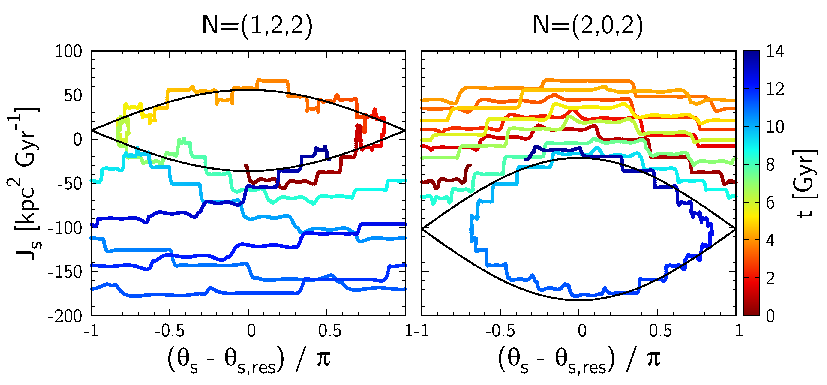}
    \vspace{-1mm}
    \caption{Both panels show the same orbit in the angle-action plane of the $\vN = (1,2,2)$ (OLR, left-hand) and the $\vN = (2,0,2)$ resonance (right-hand). Colours encode the time. The orbit is initially on the OLR and then passes to the $\vN = (2,0,2)$ resonance.}
    \label{fig:slowAA_N122vsN202}
  \end{center}
\end{figure}

To illustrate the overlapping of resonances, we show in Fig.~\ref{fig:f_res_J} the configuration of bar resonances with $\Nphi=2$ and $\Npsi=2,0,-2$ from top to bottom. A figure overlaying the three resonances is given in Appendix \ref{sec:app_overlap_of_resonances}. In a spherical system, the resonant condition only depends on $L$ and $\Jr$. However, the volume of the trapped zone depends in addition on the orbital inclination $\beta \equiv \cos^{-1}(\Lz/L)$. We thus show three plots from left to right with $\beta = 0$ (prograde), $\beta = \pi/2$ (perpendicular), and $\beta = \pi$ (retrograde). Following \cite{binney2017orbital}, we draw the furthest excursions of trapped orbits with dotted curves and mark the direction of libration with thin black lines. The colours show the phase-space density of the unperturbed distribution function (\ref{eq:app_Hernquist_DF}).

The top panels of Fig.~\ref{fig:f_res_J} show resonances with $\Npsi = 2$ which are familiar in disc dynamics: the resonance at the bottom-left corner is the corotation resonance (CR, $\Nr = 0$) and the neighbouring one is the outer Lindblad resonance (OLR, $\Nr = 1$). The trapped volume is largest at $\beta = 0$ (left, in-plane prograde), and decreases towards $\beta = \pi$ (right, in-plane retrograde) where it vanishes. The panels in the middle row present resonances with $\Npsi = 0$. Orbits trapped in these resonances complete $\Nphi$ radial oscillations in $\Nr$ bar period regardless of their azimuthal phase with respect to the bar. \cite{Weinberg2007BarHaloInteraction} called these resonances the `direct radial resonances' (DRRs). The size of DRRs peaks at $\beta = \pi/2$. The lowest panels show resonances with $\Npsi = -2$ which are strongest at $\beta=\pi$.

Comparing the different sets of resonances, we can see from Fig.~\ref{fig:f_res_J} that the resonances partly overlap, especially near $\beta \approx \pi/2$, where all resonances occupy a finite phase-space volume (see Fig.~\ref{fig:app_resonance_overlap_J} for a directly superimposed plot). To see the impact of resonance overlap, we plot in Fig.~\ref{fig:slowAA_overlap} four orbits trapped in the OLR $\vN=(1,2,2)$ with different initial actions marked by black points in Fig.~\ref{fig:f_res_J}. As expected, the two orbits at $\beta = 0$ (circle and rhombus, left column in Fig.~\ref{fig:slowAA_overlap}) are well modelled by the averaged Hamiltonian (blue), since only resonances with $\Npsi = 2$ have a non-zero volume. The right-hand column of Fig.~\ref{fig:slowAA_overlap} (triangle and square) shows two contrasting examples for $\beta = \pi/2$: the bottom right orbit (square) is mildly distorted as it is near, but still outside the second resonance, while the top right orbit (triangle) shows marked excursions as it is within the region of the $\vN=(2,0,2)$ resonance. We note that we ascribe only the large-scale distortions to the presence of the second resonance, e.g. the irregular dip near $\thetas - \thetares \approx 0$. The step-like behaviour comes instead from the high eccentricity of the orbit, i.e. its $\Js$ changes abruptly whenever the orbit rapidly passes near the bar at its pericentre.

When the domains of two resonances overlap, orbits may also pass from one resonance to another. Figure~\ref{fig:slowAA_N122vsN202} shows an example of such an orbit moving between the OLR and the $(2,0,2)$ resonance. When viewed in the slow plane of the OLR (left panel), the orbit is initially trapped and then escapes the resonance after roughly one cycle of libration. In contrast, from the viewpoint of the $(2,0,2)$ resonance (right panel), the orbit is at first outside the resonance but then becomes trapped soon after it leaves the OLR at around $9 \Gyr$.

The question here is how much does the overlap of resonances affect the calculation of the net torque based on the averaged Hamiltonian? The first thing to note is that the three sets of resonances with $\Npsi \in 2,0,-2$ have their main territories at $\beta = 0,\pi/2,\pi$ respectively, so the majority of resonant phase-space is well isolated. Secondly, the chaotic orbits found at the overlapped regions still librate quasi-periodically around one of the two resonances over a finite time interval. Thirdly, these oscillations on top of the libration behave mostly stochastically without any apparent law, so they should not leave a systematic bias when averaged over the whole set of orbits. Hence, we do not expect a significant loss of accuracy in estimating the torque using the averaged Hamiltonian. We will demonstrate this below.

\begin{figure}
  \begin{center}
    \includegraphics[width=8.5cm]{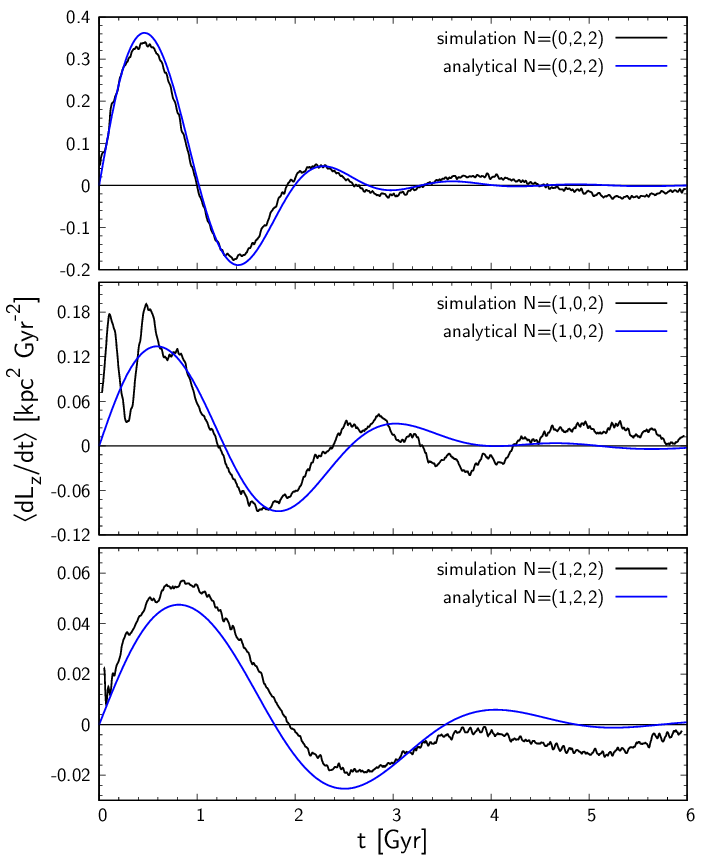}
    \vspace{-1mm}
    \caption{Total torque on orbits trapped in each resonance divided by the total halo mass. From top to bottom, the three strongest resonances are shown (note the different y-axis scale in each figure). The analytical curves (blue) agree well with that of the simulation (black).}
    \label{fig:torque_compare}
  \end{center}
\end{figure}

\subsection{Total torque on the halo}
\label{sec:total_torque_on_halo}

After these preliminary considerations, we can now achieve our goal to estimate the torque between the bar and the dark halo. Using equation (\ref{eq:torque_lib}), (\ref{eq:torque_rot}), and (\ref{eq:f2}), we obtain the total torque by summation over all resonances $\vN$:
\begin{align}
  \torq 
  &= \sum_{\vN} \int \drm^3 \vJ'' \drm^3 \vtheta'' ~f(\vtheta'',\vJ'',t) \frac{\drm \Lz}{\drm t} \nonumber \\
  &= 8(2\pi)^2 \sum_{\vN} \Nphi \int \drm^2 \vJf \Psi J_0
  \label{eq:torque_slow_regime} \\
  &\hspace{-10mm}\times \left[\int_0^1\!\drm \varepsilon \varepsilon^2 \hat{K} \!\int_0^{2\pi}\!\!\drm \thetal f(\thetal - \Omegal t,\vJ'',0)~ {\rm sn} \left(\left.\hat{K} \thetal \right\vert \varepsilon \right) {\rm dn} \left(\left.\hat{K} \thetal \right\vert \varepsilon \right) \right.\nonumber \\
  &\hspace{-10mm}\left.\pm \int_1^{\varepsilon_{\rm cut}}\!\drm \varepsilon \tilde{K} \!\int_0^{2\pi}\!\!\drm \thetac f(\thetac - \Omegac t,\vJ'',0)~ {\rm sn}\!\left(\!\left.\tilde{K} \thetac \right\vert \varepsilon^{-1} \!\right) {\rm cn}\!\left(\!\left.\tilde{K} \thetac \right\vert \varepsilon^{-1} \!\right)\right], \nonumber
\end{align}
where $f$ is the distribution function (\ref{eq:f2}) dependent on the libration/circulation angle $\thetalc$ and the resonant actions $\vJ'' = (\Jfo,\Jft,\Jlc)$, $\Psi$ is the bar perturbation (\ref{eq:app_Psi}), $J_0$ measures the width of the resonance (\ref{eq:w0_J0_eps}), and $\varepsilon$ parametrizes the value of the Hamiltonian $\bH$ (\ref{eq:w0_J0_eps}). sn, cn and dn are the Jacobi's elliptic functions (\ref{eq:app_Jacobi_ell_func}) and we have defined $\hat{K} \equiv 2 K(\varepsilon) / \pi$ and $\tilde{K} \equiv K(\varepsilon^{-1}) / \pi$ where $K$ is the complete elliptic integral of the first kind (\ref{eq:app_comp_ell_int}). The first term in the square bracket describes the torque applied on the trapped phase-space, while the second term describes the torque on the untrapped phase-space which we cut at $\varepsilon_{\rm cut} (>1)$ to avoid duplicately integrating the phase space between neighbouring resonances. However duplication cannot be fully avoided where resonances overlap. As described above and demonstrated below, we expect the effects of overlap to be minor and defer detailed modelling for the overlap regions to future study. Here, we cut the integration universally at $\varepsilon_{\rm cut} = 2-3$ which is roughly the ratio of the width and interval between the neighbouring resonances (Fig.~\ref{fig:f_res_J}). The $\pm$ sign in (\ref{eq:torque_slow_regime}) denotes integration over the upper and the lower circulating regions for which the mapping from $(\thetac,\vJ'')$ to $\vJ = (\Jr,L,\Lz)$ is different.

Figure~\ref{fig:torque_compare} compares the total torque on trapped orbits calculated with the analytical approach to a test-particle simulation. From top to bottom, we show the torques on the three strongest resonances: $(\Nr,\Npsi,\Nphi) = (0,2,2)$, $(1,0,2)$, and $(1,2,2)$. We select the trapped orbits in the simulation by demanding that they never pass $\thetas - \thetasres = \pm \pi$. We then place the particles back to their initial coordinates, re-run the simulation exclusively for these particles, and sum up their torque. For all three resonances, the analytical torque (blue) captures the principal features of the simulation (black), i.e. the torque fluctuates and damps as phase mixing progresses inside each resonance.

\begin{figure}
  \begin{center}
    \includegraphics[width=8.5cm]{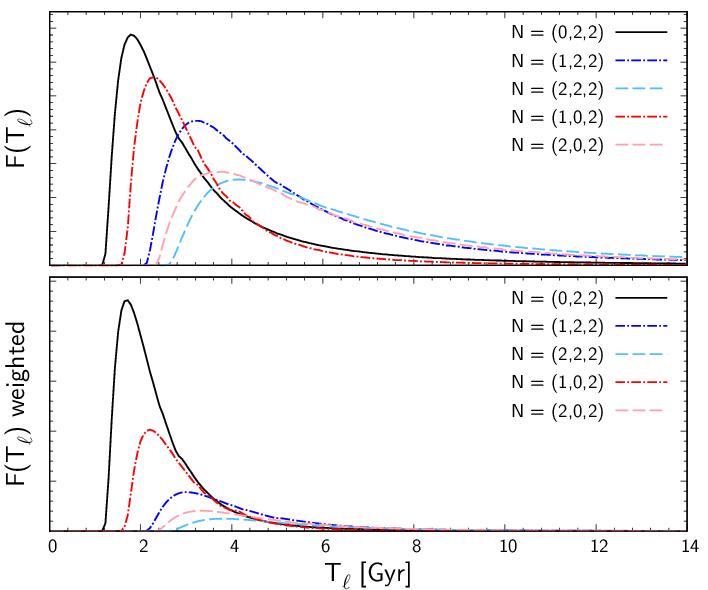}
    \vspace{-1mm}
    \caption{Top panel: Distribution of trapped orbits in libration period. Bottom panel: Distribution weighted by the amplitude of the torque which provides the actual spectrum of the total torques shown in Fig.~\ref{fig:torque_compare}.}
    \label{fig:f_Tl_combine}
  \end{center}
\end{figure}

To better quantify the time dependence of the torque, we plot the spectra of libration periods calculated analytically in the top panel of Fig.~\ref{fig:f_Tl_combine}. The distribution of each resonance peaks at a period which matches the main oscillating period of the total torque shown in Fig.~\ref{fig:torque_compare}. Since the majority of orbits with long libration periods are those trapped at small $\Psi$ (rather than those near the separatrix) and thus have little impact on the total torque, we plot on the lower panel of Fig.~\ref{fig:f_Tl_combine} the distribution when weighted by the amplitude of the torque on each orbit. Note though that this is not a Fourier analysis of Fig.~\ref{fig:torque_compare} since the torque on individual orbits becomes increasingly non-sinusoidal when approaching the separatrix (Fig.~\ref{fig:dLzdt_thetap}). The weighted distribution clarifies that the dominant contribution to the torque comes from the corotation resonance and in particular from those with libration periods $\sim 2 \Gyr$.

\begin{figure}
  \begin{center}
    \includegraphics[width=8.5cm]{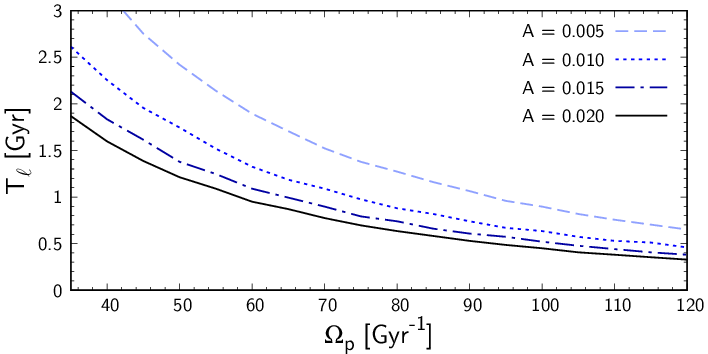}
    \vspace{-1mm}
    \caption{The libration period of the CR at the peak of its distribution $F(T_\ell)$ vs. bar pattern speed $\Omegap$ and bar strength $A$.}
    \label{fig:Tl_wp_A}
  \end{center}
\end{figure}

The libration period depends on both the bar pattern speed and the bar strength which vary over time. A simple scaling suggests $\Tl \sim \omega_0^{-1} = |G\Psi|^{-1/2} \sim \Omegap^{-1} A^{-1/2}$. Figure~\ref{fig:Tl_wp_A} shows the mode of the distribution of libration period for a given pattern speed $\Omegap$ and bar strength $A$. As expected, the libration period is shorter for higher pattern speeds and for stronger bars. Since the bar directly after formation is expected to be faster but weaker than at present, the initial torque in an expected parameter range typical for the Milky Way should have fluctuated with roughly a Gyr period.

\begin{figure}
  \begin{center}
    \includegraphics[width=8.5cm]{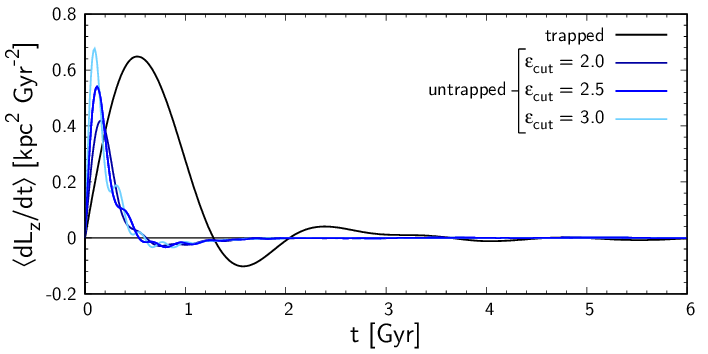}
    \vspace{-1mm}
    \caption{Comparison between the total torque exerted on resonantly trapped (black) and untrapped (blue) orbits in the dark halo calculated analytically using equation (\ref{eq:torque_slow_regime}). We cut the integration over the untrapped phase space at $\varepsilon_{\rm cut}$ to avoid duplication by neighbouring resonances. Dynamical friction by trapped and untrapped orbits rises and decays over different timescales.}
    \label{fig:torque_libvsrot}
  \end{center}
\end{figure}

Figure~\ref{fig:torque_libvsrot} compares the contribution to the total torque from trapped and untrapped phase space with three cutoff $\varepsilon_{\rm cut} = \{2,2.5,3\}$. As before, we summed over the resonances with $\Nr \in [-10,10]$, $\Npsi \in -2,0,2$, and $\Nphi = 2$. The torques on trapped and untrapped orbits rise and decay on different timescales which simply reflects the difference in their orbital periods, i.e. the period of circulation is generally shorter than that of libration. As we increase the upper limit of integration $\varepsilon_{\rm cut}$, the total torque on untrapped orbits gets larger and slightly shifts towards early time because contributions from orbits with short circulation periods far away from the resonances are added into the calculation.

\begin{figure}
  \begin{center}
    \includegraphics[width=8.5cm]{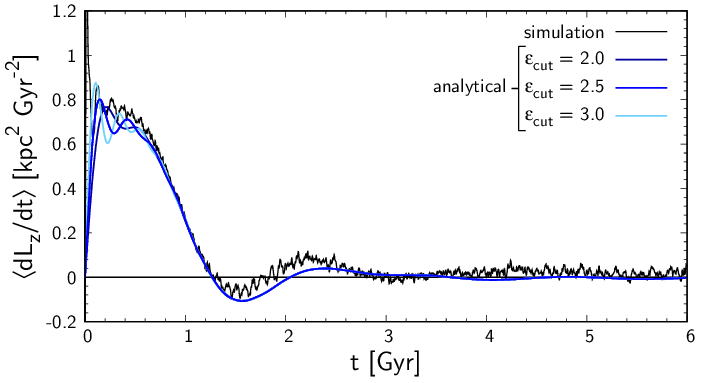}
    \vspace{-1mm}
    \caption{Total torque on the dark halo. The analytical curves (blue, equation \ref{eq:torque_slow_regime}) quantitatively explain the behaviour of the simulation (black).}
    \label{fig:torque_total}
  \end{center}
\end{figure}

Figure~\ref{fig:torque_total} shows the sum of trapped and untrapped components (blue) which quantitatively reproduces the total torque of the simulation (black). The result assures us that the trapped orbits comprise the majority of the torque at late times. In the very early times $< 0.1 \Gyr$, the torque of the simulation rises sharply as high as $1.6 {\kpc}^2 {\Gyr}^{-2}$ (see Fig.~\ref{fig:torque_W04} for the full range) but our model fails to predict this rapid response for several reasons: (i) we have neglected the fast non-resonant terms of the perturbation by averaging over the fast angles, (ii) the second order Taylor approximation makes our model inaccurate in the region of rapid circulation far from the resonances, (iii) there is a substantial volume of untrapped phase-space that were not integrated due to the cutoff at $\varepsilon_{\rm cut}$.

\subsection{Density wake in the halo}
\label{sec:density_wake_in_halo}

\begin{figure*}
  \begin{center}
      \includegraphics[width=17.8cm]{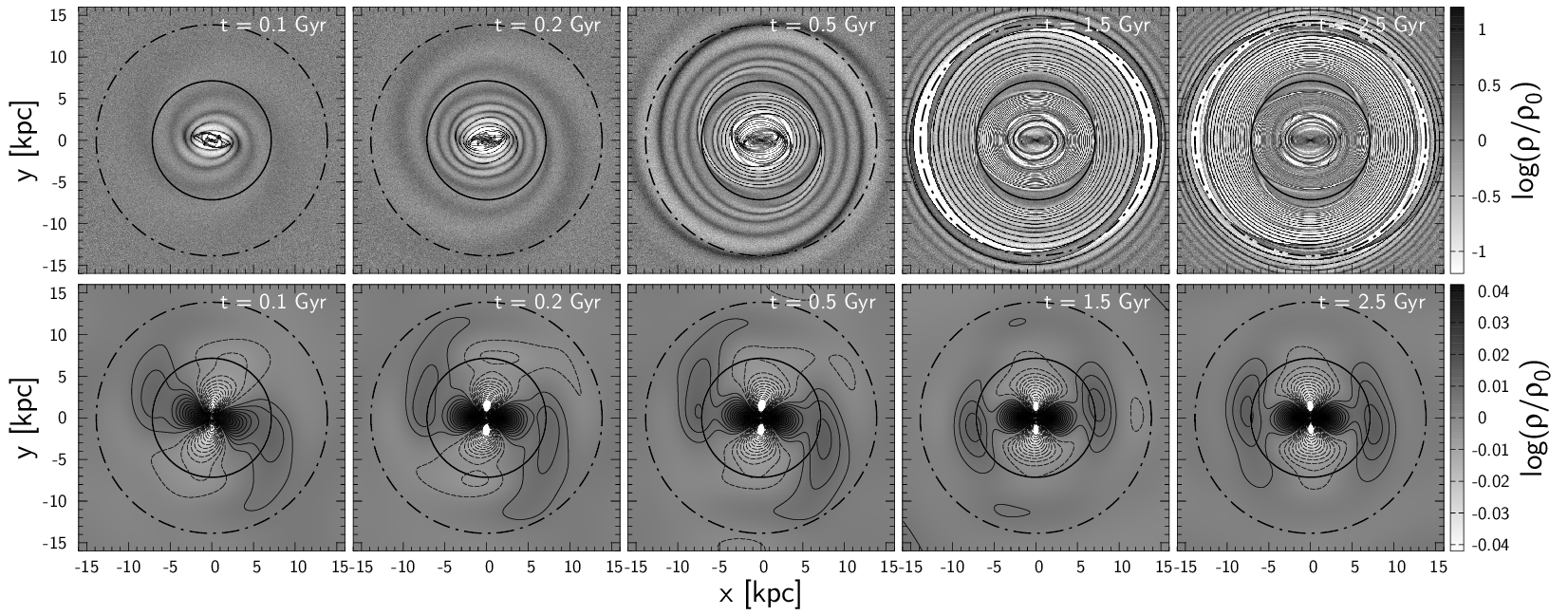}
    \vspace{-4mm}
    \caption{Density wake in the dark matter halo at the galactic mid-plane ($L=\Lz$) induced by a constantly rotating galactic bar (rotating anticlockwise). The bar lies horizontally along the $x$-axis, and we mark the radius of CR and OLR with solid and dot-dashed circles. Top row shows the density of initially circular orbits only ($\Jr=0$) and can thus be directly compared with the slow angle-action plane shown in Fig.~\ref{fig:phase_flow_slowAA}. The bottom row shows the density of all orbits at the galactic mid-plane where we have extracted the dominant pattern by representing the density with basis functions \citep{HernquistOstriker1992} restricted to azimuthal degree $l \leq 8$ and radial order $n \leq 10$. A density wake lagging behind the bar is visible in the first few hundred Myrs in line with the large positive torque on the halo during that period (Fig.~\ref{fig:torque_total}).}
    \label{fig:rho}
  \end{center}
\end{figure*}

Figure \ref{fig:rho} shows the spatial density response of the dark halo at the galactic mid-plane. The bar lies along the $x$-axis and rotates anticlockwise. The solid and dot-dashed circles represent the radii of CR and OLR respectively. The top row of Fig.~\ref{fig:rho} restricts orbits to those that are confined to the galactic mid-plane $(L=\Lz)$ and initially circular ($\Jr=0$). Their behaviour in the $x$-$y$ plane can be directly compared with the slow angle-action plane (Fig.~\ref{fig:phase_flow_slowAA}) since, for near circular orbits at the CR, the slow angle-actions are simply $(\thetas,\Js) \sim (2 \varphi, R \vc /2)$, i.e. a representation of the polar coordinates. Just after bar formation, a two-arm spiral rapidly emerges and winds up which, in Fig.~\ref{fig:phase_flow_slowAA}, corresponds to the phase-space stripes outside the separatrix. Later, a crescent-shape region becomes apparent at the CR, which corresponds to the leaf-like resonant structure in the slow angle-action plane. Vaguely, we can see an overdensity librating and phase mixing inside the resonance. The bottom row of Fig.~\ref{fig:rho} shows the density of all orbits with $L=\Lz$. By removing the restriction on $\Jr$, a variety of eccentric orbits associated with different resonances (and having different libration/circulation periods) now visit the bar region, so we no longer see a sharp signal. However, by suppressing the particle noise by representing the density with basis functions \citep{HernquistOstriker1992} restricted to finite azimuthal degree $(l \leq 8)$ and radial order $(n \leq 10)$, we can clearly identify density wakes surrounding the bar. The behaviour of the wake is consistent with the total torque on the halo (Fig.~\ref{fig:torque_total}): In the first few 100 Myrs (first three panels), a prominent density wake lags behind the bar, resulting in a large positive torque on the halo (negative torque on the bar). At around $1.5 \Gyr$, the wake moves to the bar front, hence receiving a net negative torque. Subsequently, the wake converges to a symmetric configuration around the bar, and the net torque diminishes.

\section{Slowdown of the bar}
\label{sec:slowdown_bar}

The natural progression of this work will be a self-consistent slowdown model of the bar. However, the bar's moment of inertia is not a constant, the halo's self-gravity will amplify the torque, and the inflow of gas towards the nuclear disc provides an additional angular momentum term. Resolving these issues is beyond the scope of this paper, and so must be deferred to future analysis. Instead, here we will discuss the effects of bar slowdown on the angular momentum balance using the bar's slowing rate directly measured from local stellar kinematics \citep{Chiba2020ResonanceSweeping} which agrees with \textit{N}-body+SPH simulation of the bar in a live dark halo \citep[][]{aumer2015origin}.

\begin{figure*}
  \begin{center}
    \setlength\columnsep{-40pt}
    \begin{multicols}{2}
      \includegraphics[width=8.0cm]{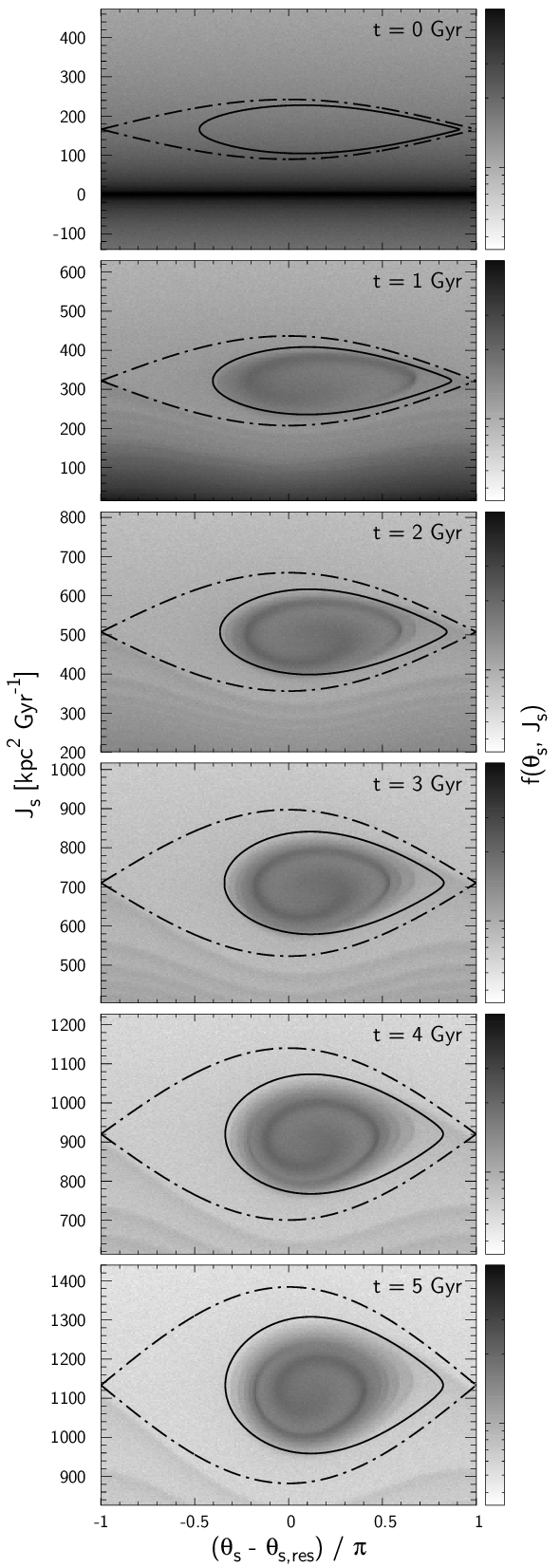}
      \newpage
      \includegraphics[width=8.3cm]{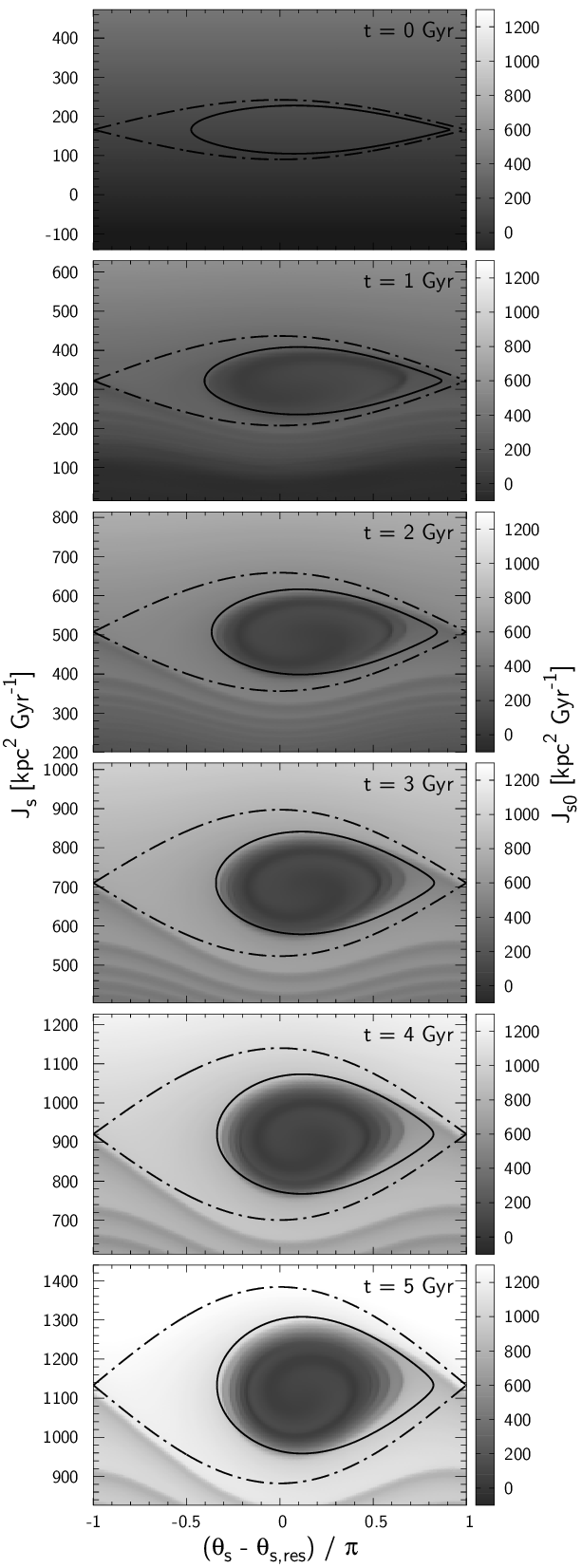}
    \end{multicols}
    \vspace{-6mm}
    \caption{Time evolution of the phase-space near the CR when the bar is slowing down. The left column shows the phase-space density in log scale, while the right column shows the initial $\Js$. The phase-space spiral continues to wind up as the resonance migrates upwards (note the increasing $y$ axis from the top panel to the bottom). The bar slows down at rate $\eta = - \dOmegap/\Omegap^2 = 0.004$ starting from $\Omegap = 81.8 \iGyr$ and ends up being $\Omegap = 31.0 \iGyr$ after $5 \Gyr$.}
    \label{fig:phase_flow_slowAA_slow}
  \end{center}
\end{figure*}

Figure~\ref{fig:phase_flow_slowAA_slow} shows the phase space perturbed by a bar that slows with $\Omegap \propto 1/t$ from an initial value $\Omegapo = 81.8 \iGyr$ with rate $\eta \equiv -\dOmegap/\Omegap^2 = 0.004$\footnote{A note on wording: This slowing rate implies $s \sim \eta/A = 0.2$ which is in the `slow' regime $(s < 1)$. However, \cite{Chiba2020ResonanceSweeping} classified this model as a `rapidly' decelerating bar in contrast to a yet slower model that evolves almost adiabatically.}. The bar strength is fixed to $A=0.02$. As the bar slows, the resonantly trapped orbits get dragged with the resonance towards larger $\Js$ while they keep phase mixing internally. The dot-dashed curves show the separatrix evaluated in a `time-frozen' Hamiltonian (when fixing the pattern speed) while the solid curves mark the approximate boundary of orbits that remain trapped in the moving resonance (Appendix \ref{sec:app_orbit_subject_to_slowing_bar}). As described in \cite{Chiba2020ResonanceSweeping}, the deceleration tilts the potential of the resonance (\ref{eq:app_Ep}), making the effective trapped region shrink and shift towards positive $\thetas$. The latter effect could also be explained by the Euler force arising from the deceleration of the rotating frame of reference.

\subsection{Capture and escape from resonance}
\label{sec:resonant_capturing}

Resonance sweeping leaves a significant change in the distribution through capture and loss. When the resonance sweeps outwards, the untrapped orbits swept over by the upper separatrix (dot-dashed curves in Fig.~\ref{fig:phase_flow_slowAA_slow}) will temporarily get trapped, as adiabaticity is broken near the separatrix. Orbits that crossed the upper separatrix near $\thetas - \thetares = \pi$ may reach the adiabatic region (solid curves), where the libration action is conserved, and thus remain trapped and get dragged by the resonance. Other orbits will escape the resonance before completing one cycle of libration, proceeding to the lower circulating zone. Depending on the angle $\thetas$ at which these uncaptured orbits enter the resonance, they will travel a different amount of distance in $\Js$: orbits crossing the separatrix near $\thetas - \thetares = - \pi$ will immediately escape the resonance, ending up at a relatively low $\Js$, while those that could almost but not quite reach the adiabatic region will be dragged a large distance in $\Js$ before they get damped out of the resonance. Thus, after the passage of the resonance, orbits originally at the same action $J_{\rm s0}$ (same colour in the right column of Fig.~\ref{fig:phase_flow_slowAA_slow}) will spread over a wide range of $\Js$. These orbits then phase mix and form phase-space stripes in the lower circulating region.

As seen in Fig.~\ref{fig:phase_flow_slowAA_slow}, the entire trapped volume (and the adiabatic region within) generally grows as the bar slows. Inside the adiabatic region, trapped orbits conserve their $\Jl$, so the newly captured orbits accrete on the surface of the adiabatic region. As a result, the resonance grows like a tree ring, where the $\Js$ of origin increases from the core towards the surface \citep{Chiba2021TreeRing}.

Previous studies formalized the capturing and escaping processes in the adiabatic limit ($s \ll 1$) by the transition probability averaged over the libration/circulation angles \citep[e.g.][]{Henrard1982Capture,Sridhar1996Adiabatic,Collett1997Capture}. However, to model the transient behaviour of dynamical friction, a phase dependent prescription for capture and loss is required. A straightforward way would be to numerically integrate the equations of motion for the averaged system with time-dependent parameters \citep{Weinberg2007BarHaloInteraction}. Alternatively, one could return to the Eulerian approach (section \ref{sec:torque_fast_regime}) and fully address the non-linear terms of the CBE. This approach can in principle model the resonant structure with arbitrary time-dependence (e.g. bar growth), although it has the difficulty that increasing numbers of higher-order modes will be required to model the fine structures developed by phase mixing.

\subsection{Transfer of angular momentum by resonant dragging}
\label{sec:resonant_dragging}

In addition to the torque caused by phase mixing, the moving resonances give rise to another type of angular momentum transfer. The conservation of libration action ensures that the adiabatic region preserves the phase-space density at the time of capture. Since the resonances move towards higher angular momentum, this implies that the population on resonances is much denser than the surrounding phase space. When resonances move outward, the freed phase space below them will be filled with orbits jumping across the resonances, and just like lifting a heavy body in water, the halo gains net angular momentum which further slows the bar (i.e. positive feedback).

If we accelerate the bar instead, the resonances sweep inwards. As long as they are still denser than the surrounding phase space, the resonance frees angular momentum, further accelerating the bar (i.e. again, positive feedback). However, as the resonances proceed deep into the inner halo, the surrounding density rises and so the amount of freed angular momentum decreases up to even becoming negative when the resonances are underdense (i.e. negative feedback).

\citetalias{Tremaine1984Dynamical} and \cite{weinberg1985evolution} called these types of angular momentum exchange the `dynamical feedback' and presented a formula to predict its amount, although they only figured in the torque due to orbits that jump across the resonance and separately discussed the effect of resonant dragging. They gave an insightful interpretation of dynamical feedback as a change in the moment of inertia of the bar: since dynamical feedback is to first order proportional to the change in pattern speed, ($\tau_{\rm feedback} = C \dOmegap,$ where $C>0$), the equation of motion for the bar becomes \citep{weinberg1985evolution}
\begin{align}
  \dOmegap = \frac{\tau_{\rm friction} + \tau_{\rm feedback}}{\Ib} = \frac{\tau_{\rm friction}}{\Ib - C}
  \label{eq:effective_moment_of_inertia}
\end{align}
where $\tau_{\rm friction}$ represents dynamical friction and $\Ib$ is the bar's moment of inertia. This equation implies that dynamical feedback translates as a reduction in the moment of inertia. However we caution that (i) dynamical feedback is always delayed by an order of the libration period, meaning that for a rapidly fluctuating torque caused by e.g. the bar-spiral interactions, the bar will react with its bared moment of inertia. (ii) $C$ is not a constant and depends both on the pattern speed $\Omegap$ as well as its derivative $\dOmegap$. In particular, $C$ is a decreasing function of $\dOmegap$ (i.e. the faster the bar slows, the smaller the adiabatic region) which implies that dynamical feedback stabilizes the bar's slowing rate.

\section{Summary}
\label{sec:conclusions}

This work is a first attempt to push beyond the standard theory of dynamical friction between galactic bars (or in fact any galactic substructures) and dark matter haloes in the \textit{slow regime}, i.e. for bars decelerating slowly enough to allow resonant trapping as observed in the Milky Way. In particular, we have:
\begin{itemize}
\item shown how the standard linear perturbation theory of dynamical friction breaks down in the slow regime due to the growth of non-linear responses near resonances,
\item developed an improved analytical description for angular momentum transfer in the slow limit using the resonant angle-action coordinates and demonstrated numerically that the same mechanism applies generally to a growing or slowing bar,
\item shown that dynamical friction oscillates with the libration period of the main resonances, which may drive a significant fluctuation in bar pattern speed at early times.
\end{itemize}

As first discussed in the seminal paper by \citetalias{Tremaine1984Dynamical}, the dynamic response to a perturbation qualitatively changes depending on the speed of resonance that sweeps the phase space. This is characterized by the parameter $s = |\nphi \dOmegap / \omega_0^2|$ that measures the rate of change in pattern speed $\dOmegap$ with respect to the libration frequency $\omega_0$; the regime is called \textit{fast} when $s > 1$ and \textit{slow} otherwise. While earlier simulations have reported bar evolution in the fast regime \citep[e.g.][]{weinberg1985evolution,hernquist1992bar,Sellwood2006BarHaloI}, recent observation of the Milky Way disc \citep{Chiba2020ResonanceSweeping} as well as $N$-body+hydrodynamic simulations of a realistically growing disc \citep{aumer2015origin} indicate that the main bar resonances evolve in the slow regime, trapping a host of dark matter and disc stars in resonance. We note however that, while some resonances evolve in the slow regime, others may simultaneously evolve in the fast regime since $\omega_0$ varies from resonance to resonance (and even within a single resonance, e.g. depending on the orbital inclination). Therefore, fast and slow regimes are both important in bar evolution.

\citetalias{Tremaine1984Dynamical} and \citetalias{Weinberg2004Timedependent} have shown that bar evolutions in the fast limit $(s \gg 1)$ can be well modelled by linear perturbation theory because the resonances will sweep past the orbits before any non-linearity can grow. In the slow limit $(s \rightarrow 0)$, however, the linear assumption only holds for a fraction of the libration period (Fig.~\ref{fig:torque_W04}, Appendix \ref{sec:app_linear_response}). \citetalias{Tremaine1984Dynamical} has provided a prescription for arbitrary $s$ by Taylor approximation about the resonance although this approach cannot model the transient responses (phase mixing), which may last for several Gyrs.

We have developed a fully time-dependent theory of dynamical friction in the slow limit by modelling orbits near resonances with the resonant angle-action coordinates. In these new coordinates, the resonant motions become perfectly linear: trapped/untrapped orbits evolve linearly in their libration/circulation angles at constant rates.

The resonantly trapped orbits periodically exchange $z$-angular momentum with the bar as they slowly librate around the resonance. For instance, orbits trapped at the corotation resonance around the stable Lagrange points have increasing angular momentum while on the trailing side of the bar and are losing angular momentum while on the leading side. The net transfer of angular momentum is determined by the libration angle distribution: For a typical halo in equilibrium, the freshly trapped orbits are predominantly at lower angular momentum (libration angle near $\thetal \sim \pi$). This phase imbalance initially leads to a net negative torque on the bar as this overdensity collectively librates up to higher angular momentum. However, after half a libration period, the overdensity starts librating back to lower angular momentum and can transfer angular momentum back to the galactic bar. Since libration periods increase towards the separatrix, this overdensity gradually winds up into a phase-space spiral inside the resonance and the net torque follows secular damped oscillations. The untrapped orbits circulating outside the separatrix similarly exchange angular momentum with the bar as they phase mix, although they have a much shorter mixing time.

We have shown that due to the near-harmonic effective potential around the resonance centre, libration periods are quite similar across the trapped volume. This provides enough coherence to cause several oscillations in the torque. A straightforward prediction from this is that after bar formation, the pattern speed of the bar fluctuates with the typical libration frequencies of the main resonances. And indeed, while we are not aware of this having been noticed in previous literature, a closer inspection of previous \textit{N}-body simulations shows these predicted pattern speed oscillations \citep[e.g.][]{aumer2015origin}. We note that these oscillations in bar pattern speed take place on an order of magnitude longer timescale $(>1 \Gyr)$ than the short-term oscillations caused by alignments between bar and spiral patterns \citep{Wu2016TimeDependent,Hilmi2020Fluctuations}.

The resonant angle-action coordinates for each resonance are constructed by averaging the Hamiltonian over the fast angles. The averaged Hamiltonian is, however, inaccurate when neighbouring resonances lie too close: overlap of resonances leads to the onset of chaos where stochastically orbits move from one resonance to another in a stochastic manner. In our current bar model, the limited regions of chaotic phase-space did not significantly affect the estimation of the total torque, although this may not be the case with a more realistic bar with additional higher order modes (e.g. octopole). 

This paper has shown qualitatively how dynamical friction may work in the slow regime. To obtain quantitative predictions, a self-consistent model is required. In particular, we must model (i) the growth of the bar, which changes its effective moment of inertia, (ii) the self-gravitational perturbation of the halo, which can amplify the torque on the bar by several factors \citep[e.g.][]{Weinberg1989SelfGravitating,Chavanis2012Kinetic,Dootson2022}, and (iii) the angular momentum supply from the gas, which sheds angular momentum to the bar as it sinks towards the Galactic centre \citep[e.g.][]{Regan2004BarDrivenMassInflow,berentzen2007gas,Athanassoula2013BarGas}.

Finally, the phase-space spiral predicted in this paper is also expected to form inside the resonances of the stellar disc. If we can identify this spiral pattern in the observational data, it will allow us to directly constrain the age of the Galactic bar.

\section*{Acknowledgements}

It is our pleasure to thank W. Dehnen for providing us with detailed comments on our draft. We also thank J. Binney, J. Magorrian, C. Hamilton, E. Athanassoula, and M. Semczuk for many helpful comments. We particularly thank M. Weinberg for a fruitful and enlightening discussion as well as helping us greatly improve the paper as a referee. R.C. acknowledges financial support from the Takenaka Scholarship Foundation and the Royal Society grant RGF$\backslash$R1$\backslash$180095. R.S. is supported by a Royal Society University Research Fellowship. This work was performed using the Cambridge Service for Data Driven Discovery (CSD3), part of which is operated by the University of Cambridge Research Computing on behalf of the STFC DiRAC HPC Facility (www.dirac.ac.uk). The DiRAC component of CSD3 was funded by BEIS capital funding via STFC capital grants ST/P002307/1 and ST/R002452/1 and STFC operations grant ST/R00689X/1. DiRAC is part of the National e-Infrastructure.

\section*{Data availability}

The codes used to produce the results are available from the corresponding author upon request.



\bibliographystyle{mnras}
\bibliography{./references}



\appendix

\section{Distribution function of Hernquist halo}
\label{sec:app_DF}

The distribution function of an isotropic Hernquist halo normalized to 1 is \citep{Hernquist1990AnalyticalModel}
\begin{align}
  f(E) &= \frac{3\sin^{-1}\sqrt{\varE} + \sqrt{\varE(1-\varE)}(1-2\varE)(8\varE^2-8\varE-3)}{8\sqrt{2}\pi^3\rs^3\vg^3 (1 - \varE)^{\frac{5}{2}}}
  \label{eq:app_Hernquist_DF}
\end{align}
where $\vg = \sqrt{G M /\rs}$, and $\mathcal{E} = - E/\vg^2$ is the dimensionless binding energy. We draw initial condition for the test particles by first sampling the energy from the differential energy distribution \citep{binney2008galactic}
\begin{align}
  \mathcal{N}(E) &= f(E) g(E)
  \label{eq:app_Hernquist_dNdE}
\end{align}
where $g(E)$ is the density of states (i.e. phase-space volume per unit energy)
\begin{align}
  g(E) &= \frac{2\sqrt{2}\pi^2\rs^3\vg}{3\varE^{\frac{5}{2}}} \bigl[ 3(8\varE^2 - 4\varE + 1) \cos^{-1}\sqrt{\varE} \nonumber \\ 
  &\hspace{30mm} - \sqrt{\varE(1 - \varE)} (4\varE - 1)(2\varE +3) \bigr].
  \label{eq:app_Hernquist_g}
\end{align}
We then choose the initial radius $r$ at fixed $E$ from the distribution 
\begin{align}
  P(r) &\propto \int \drm^3 \vx \drm^3 \vvel \delta\left[\frac{1}{2} \vvel^2 + \Phi(\vx) - E \right] \delta\left[|\vx| - r\right] \nonumber \\
  &= (4 \pi r)^2 \sqrt{2\left[E-\Phi(r)\right]}.
  \label{eq:app_Hernquist_Pr}
\end{align}
With $E$ and $r$, the initial speed $v = \sqrt{2\left[E-\Phi(r)\right]}$ is determined. Finally, we pick angles randomly over a sphere in position space (i.e. spherical) and in velocity space (i.e. isotropic). We confirmed that the density and anisotropy distribution of test particles constructed in this manner are unaltered after $10 \Gyr$ of iteration without the bar.

\section{Fourier coefficients of bar potential}
\label{sec:app_Psi}

We follow the standard method of \citetalias{Tremaine1984Dynamical}, first expanding the bar's potential in spherical harmonics
\begin{align}
  \Phib(r,\vartheta,\varphi,t) = \sum_{l = 0}^{\infty} \sum_{m = - l}^{l} \Philm(r) Y_{lm}\left(\vartheta,\varphi - \Omegap t\right).
  \label{eq:app_bar_potential_sperical_harmonics}
\end{align}
For our bar model (\ref{eq:bar_potential}), only terms with $l = 2$ and $m = \pm 2$ are non-zero and their coefficients are
\begin{align}
  \Phi_{2,2}(r) = \Phi_{2,-2}(r) = \frac{\Phib(r)}{2 Y_{2,2}(\pi/2,0)},
  \label{eq:app_bar_potential_Philm}
\end{align}
where $\Phib(r)$ is the radial profile of the bar potential (\ref{eq:bar_potential_amp}). The Fourier coefficients $\hPhin(\vJ,t)$ of the bar potential expanded in the angles $\vtheta = (\thetar,\thetapsi,\thetaphi)$ can then be expressed as \citepalias{Tremaine1984Dynamical}
\begin{align}
  &\hPhin(\vJ,t) = \int \frac{\drm^3 \vtheta}{(2 \pi)^3} \Phib(r,\vartheta,\varphi,t) \e^{-i \vn \cdot \vtheta} \nonumber \\
  &= \sum_{l = 0}^{\infty} \sum_{m = - l}^{l} i^{m - \npsi} \delta_m^{\nphi} Y_l^{\npsi}\left(\frac{\pi}{2},0\right) d_{\npsi m}^{l}(\beta) W_{lm}^{\nr\npsi}(\vJ) \e^{-i m \Omegap t},
  \label{eq:app_bar_potential_Fourier}
\end{align}
where $d_{\npsi m}^{l}(\beta)$ is the Wigner's small $d$-matrix \citep[e.g.][]{Wigner1959Group}
\begin{align}
  &d_{\npsi m}^{l}(\beta) = \sum_t (-1)^t \frac{\sqrt{(l+\npsi)!(l-\npsi)!(l+m)!(l-m)!}}{(l-m-t)!(l+\npsi-t)!t!(t+m-\npsi)!} \nonumber \\
  &\hspace{2.5cm} \times \left(\cos\frac{\beta}{2}\right)^{2l+\npsi-m-2t} \left(\sin\frac{\beta}{2}\right)^{2t+m-\npsi}.
  \label{eq:app_Wigner_d_Matrix}
\end{align}
and
\begin{align}
  &W_{lm}^{\nr\npsi}(\vJ) \equiv \frac{1}{\pi} \int_0^{\pi} \drm \thetar \Philm(r) \cos\left[\nr \thetar + \npsi (\thetapsi - \psi)\right].
  \label{eq:app_bar_potential_W}
\end{align}
Note that $\thetapsi - \psi$ (the azimuthal deviation from the guiding centre) only depends on $\thetar$ for a given $\vJ$, so the integral leaves no dependence on the angles. We may exert the $\delta_m^{\nphi}$ in (\ref{eq:app_bar_potential_Fourier}) and express the coefficients in the form of $\hPhin(\vJ,t) = \hPhin(\vJ) \e^{-i \nphi \Omegap t}$ where
\begin{align}
  \hPhin(\vJ) \equiv &\sum_{l = 0}^{\infty} i^{\nphi - \npsi} Y_l^{\npsi}\left(\frac{\pi}{2},0\right) d_{\npsi \nphi}^{l}(\beta) W_{l\nphi}^{\nr\npsi}(\vJ).
  \label{eq:app_bar_potential_PhinJ}
\end{align}

The Fourier coefficients of the bar potential expanded in the slow-fast angles $\vtheta' = (\thetafo,\thetaft,\thetas)$ at resonance $\vN$ are
\begin{align}
  \hPsik(\vJ')
  &= \int \frac{\drm^3 \vtheta'}{(2 \pi)^3} \Phib(r,\vartheta,\varphi,t) \e^{-i \vk \cdot \vtheta'} \nonumber \\
  &= \int \frac{\drm^3 \vtheta'}{(2 \pi)^3} \left[\sum_{\vn} \hPhin(\vJ) \e^{i \left(\vn \cdot \vtheta - \nphi \Omegap t\right)} \right] \e^{-i \vk \cdot \vtheta'} \nonumber \\
  &= \sum_{\vn} \hPhin(\vJ) \int \frac{\drm^3 \vtheta'}{(2 \pi)^3} \nonumber \\
  &\hspace{-0.4cm}\times\e^{i \left[\left(\nr - \Nr \frac{\nphi}{\Nphi} - \kfo\right)\thetafo + \left(\npsi - \Npsi \frac{\nphi}{\Nphi} - \kft\right)\thetaft + \left(\frac{\nphi}{\Nphi} - \ks\right)\thetas \right]},
\end{align}
where the last line follows from the relation $(\thetar,\thetapsi,\thetaphi) = (\thetafo,\thetaft,(\thetas-\Nr\thetafo-\Npsi\thetaft)/\Nphi+\Omegap t)$. Since $\nphi$ is restricted to $m = \pm 2$, we have for resonances with $\Nphi = 2$
\begin{align}
  \hPsik(\vJ') &= \sum_{\vn} \hPhin(\vJ) \int \frac{\drm^3 \vtheta'}{(2 \pi)^3} \nonumber \\
  &\hspace{15mm} \times \e^{i \left[\left(\nr \mp \Nr - \kfo\right)\thetafo + \left(\npsi \mp \Npsi - \kft\right)\thetaft + \left(\pm 1 - \ks\right)\thetas \right]} \nonumber \\
  &= \sum_{\vn} \hPhin(\vJ) \delta_{\nr, \kfo \pm \Nr} \delta_{\npsi, \kft \pm \Npsi} \delta_{\ks ,\pm 1} \nonumber \\
  &= \hPhi_{(\kfo \pm \Nr, \kft \pm \Npsi, \pm \Nphi)}(\vJ) \delta_{\ks ,\pm 1}.
  \label{eq:app_bar_potential_Fourier_slowfast}
\end{align}
Therefore, the resonant term $\Psi(\vJ') \equiv 2 \left\vert \hPsi_{(0,0,1)}(\vJ') \right\vert$ in the averaged Hamiltonian (\ref{eq:Hamiltonian}) is
\begin{align}
  \Psi(\vJ') = 2 \left\vert \sum_{l = 0}^{\infty} i^{\Nphi - \Npsi} Y_l^{\Npsi}\left(\frac{\pi}{2},0\right) d_{\Npsi \Nphi}^{l}(\beta) W_{l\Nphi}^{\Nr\Npsi}(\vJ) \right\vert.
\label{eq:app_Psi}
\end{align}

\section{Conservation of circulation action}
\label{sec:app_conservation_Jc}

\begin{figure}
  \begin{center}
    \includegraphics[width=8.5cm]{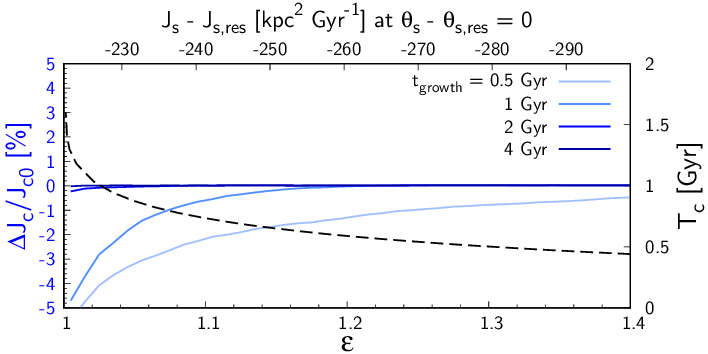}
    \vspace{-1mm}
    \caption{Changes in circulation action $\Jc$ (blue, left $y$-axis) at the CR as a function of $\varepsilon$ and bar growth time $t_{\rm growth}$. The black dashed curve shows the circulation period (right $y$-axis).}
    \label{fig:Jc}
  \end{center}
\end{figure}

Figure~\ref{fig:Jc} demonstrates the conservation of circulation action $\Jc$ in our growing bar simulation presented in Fig.~\ref{fig:phase_flow_slowAA_grow}. The blue curves show the change in circulation action $\Delta \Jc / J_{\rm c0} = (\Jc - J_{\rm c0})/J_{\rm c0}$ of untrapped orbits in the lower circulating region for a given bar growth time $t_{\rm growth}$. $J_{\rm c0}$ and $\Jc$ are obtained by numerically integrating the orbits' $\Js$ over $\thetas$ (\ref{eq:Jc}) before and after bar formation. We plot the mean change in $\Jc$ averaged over the final circulation angle $\thetac$. The $x$-axis ranges from $\varepsilon = 1$ (separatrix) to $1.4$ and the top $x$-axis indicates the corresponding coordinate in $\Js-\Jsres$ at $\thetas-\thetasres = 0$. The change in $\Jc$ becomes larger towards the separatrix and with shorter bar-growth time. With $t_{\rm growth} = 4$ Gyr used in Fig.~\ref{fig:phase_flow_slowAA_grow}, most of the untrapped orbits conserve $\Jc$ to high accuracy $(<0.1\%)$. For reference, we also plot the circulation period calculated by orbit integration (black dashed, right $y$-axis) which is several factors smaller than the adopted $t_{\rm growth}$ apart from the vicinity of the separatrix.

\section{Linear response}
\label{sec:app_linear_response}

As discussed in \citetalias{Tremaine1984Dynamical}, the linear response remains valid so long as the bar's pattern speed changes sufficiently rapidly (the fast limit, $s \gg 1$). In the slow regime ($s < 1$), however, the error of linear approximation increases due to the growth of non-linearity near resonances. In section \ref{sec:torque_fast_regime}, we have demonstrated this in the slow limit ($s \rightarrow 0$), showing that linear theory fails to predict the torque beyond few hundred Myrs. Here we take a closer look at this issue by directly observing the distribution function in the slow angle-action plane. The Fourier coefficients of the linear response (\ref{eq:fn_linearizedCBE}) to a constantly rotating perturbation is
\begin{align}
  \hfn(\vJ, t) 
  &= i \vn \cdot \frac{\pd f_0}{\pd \vJ} \int_0^t \drm t' \e^{- i \vn \cdot \vOmega (t - t')} \left[ \hPhin(\vJ) \e^{- \nphi \Omegap t'} \right] \nonumber \\
  &= i \vn \cdot \frac{\pd f_0}{\pd \vJ} \hPhin(\vJ) \e^{- i \vn \cdot \vOmega t} \left[ \int_0^t \drm t' \e^{i \left(\vn \cdot \vOmega - \nphi \Omegap \right) t'} \right] \nonumber \\
  &= \vn \cdot \frac{\pd f_0}{\pd \vJ} \hPhin(\vJ) \frac{ \e^{- i \nphi \Omegap t} - \e^{- i \vn \cdot \vOmega t}}{\vn \cdot \vOmega - \nphi \Omegap}.
  \label{eq:app_fn_linearizedCBE}
\end{align}
The total linear response is then
\begin{align}
  f_1(\vtheta, \vJ, t) 
  &= \sum_{\vn} \hfn(\vJ, t) \e^{i \vn \cdot \vtheta} \nonumber \\
  &= \sum_{\vn} \vn \cdot \frac{\pd f_0}{\pd \vJ} \hPhin(\vJ) \frac{ \e^{i \left( \vn \cdot \vtheta - \nphi \Omegap t \right)} - \e^{i \left(\vn \cdot \vtheta - \vn \cdot \vOmega t \right)}}{\vn \cdot \vOmega - \nphi \Omegap}.
  \label{eq:app_linear_response}
\end{align}
Note that this equation contains the resonant frequency at the denominator but it does not diverge at the resonance because the numerator also vanishes there. From the condition that the perturbed potential is real $\hat{\Phi}_{-\vn}(\vJ,t) = \hat{\Phi}^{\ast}_{\vn}(\vJ,t)$, one can express $f_1$ as
\begin{align}
  f_1(\vtheta, \vJ, t) 
  &\!=\! \sum_{\vn} \vn \!\cdot\! \frac{\pd f_0}{\pd \vJ} \left\vert\hPhin(\vJ)\right\vert \frac{\cos{\left(\thetas \!-\! \thetasres\right)} \!-\! \cos{\left(\thetas \!-\! \thetasres \!-\! \Omegas t\right)}}{\Omegas} \nonumber \\
  &\!=\! -\sum_{\vn} \vn \!\cdot\! \frac{\pd f_0}{\pd \vJ} \left\vert\hPhin(\vJ)\right\vert \sin\left(\thetas \!-\! \thetasres \!-\! \Omegas t/2\right) \frac{\sin\left(\Omegas t / 2\right)}{\Omegas / 2},
  \label{eq:app_linear_response_real}
\end{align}
where $\thetas = \vn \cdot \vtheta - \nphi \Omegap t$ and $\Omegas = \vn \cdot \vOmega - \nphi \Omegap$. A similar equation is presented in \cite{Weinberg2007BarHaloInteraction}. At the resonance $\Omegas \rightarrow 0$, the term $\sin \left(\Omegas t / 2\right) / \left(\Omegas / 2\right)$ becomes $t$, implying linear growth of perturbation $f_1$ in time. For linear theory to hold, $f_1$ must remain sufficiently smaller than the variation of the background distribution $\Delta f_0$ across the resonance $\Delta \Js \sim J_0 = \sqrt{-\Psi/G}$ (\ref{eq:w0_J0_eps}). This requires
\begin{align}
  1 \gg \frac{f_1}{\Delta f_0} \sim \frac{\Psi}{\Delta \Js} t \sim \sqrt{-\Psi G} t = \omega_0 t.
  \label{eq:app_linear_timescale}
\end{align}
Hence the timescale at which linear theory is valid in the slow limit is set by the libration period $\omega_0^{-1} \gg t$.

\begin{figure*}
  \begin{center}
      \includegraphics[width=17.8cm]{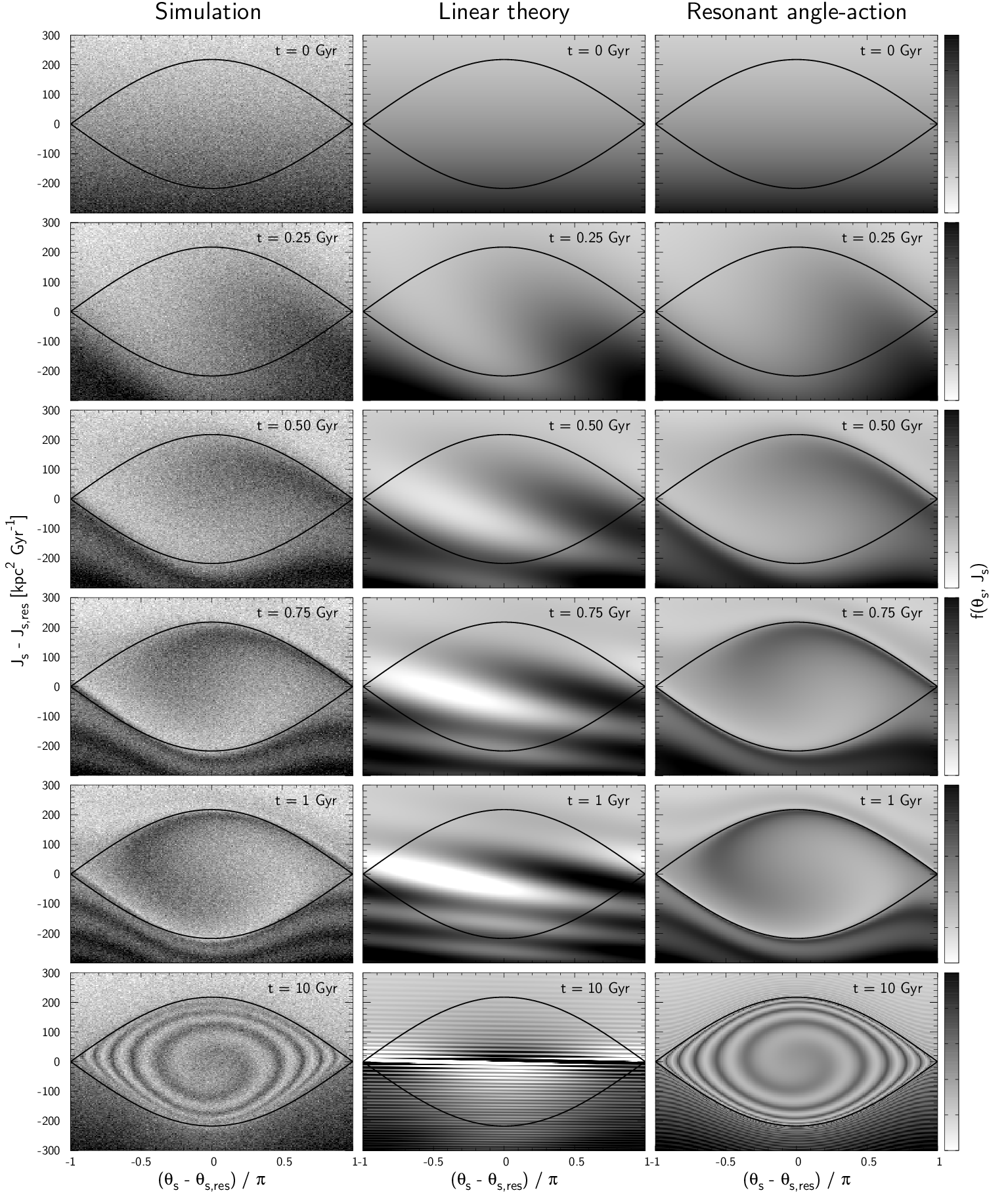}
    \vspace{-0mm}
    \caption{Time evolution of the phase space density near the corotation resonance at $(\Jfo,\Jft)=(10,0) \kpckpcGyr$. Comparison between simulation (left), linear perturbation theory (middle) and resonant angle-action theory (right). Linear theory properly predicts the phase flow for the first few hundred Myrs but fails to predict the subsequent behaviour.}
    \label{fig:linear_vs_secular}
  \end{center}
\end{figure*}

Figure~\ref{fig:linear_vs_secular} compares the simulated phase flow near the corotation resonance (left) with linear theory (middle) and the resonant angle-action theory (right) described in section \ref{sec:torque_slow_regime}. Here, the libration period at the core of the resonance is $T_\ell = 2\pi/\omega_0 \sim 1.2 \Gyr$. For the top five plots, the time since bar formation increases in steps of $0.25 \Gyr$, while the bottom plot shows the result after $10 \Gyr$. The linear response oscillates sinusoidally in $\thetas$ with phase velocities that vanish towards the resonance, as described by the term $\sin{\left(\thetas \!-\! \thetasres \!-\! \Omegas t/2\right)}$ in (\ref{eq:app_linear_response_real}). For the first few hundred Myrs (top two rows), this shearing motion results in a change in distribution similar to that caused by the motion of libration and circulation. However, errors in linear theory become prominent beyond $t \gtrsim 0.2 T_\ell$.

Despite the growing error in linear theory, past works (e.g. \citetalias{Tremaine1984Dynamical}; \citetalias{lynden1972generating}) have commonly assumed that the perturbation was switched on in the distant past which is equivalent to taking $t \rightarrow \infty$. As noted in \cite{Weinberg2007BarHaloInteraction}, in this time-asymptotic limit, the function $\sin \left(\Omegas t / 2\right) / \left(\Omegas / 2\right)$ in (\ref{eq:app_linear_response_real}) approaches a delta function $\pi\delta(\Omegas/2)$\footnote{
In the sense that,
\begin{align}
  \lim_{t \rightarrow \infty} \int \drm^3 \vJ g(\vJ) \frac{\sin \left(\Omegas t / 2\right)}{\Omegas / 2} = g(\vJ) \pi \delta(\Omegas/2)
\end{align}
}, i.e. the wavelength of the linear response in frequency (or action) space gets narrower and narrower with time, and eventually an integration over phase space will only leave contribution from the perfectly resonant orbits. This is the prediction of the LBK formula (\ref{eq:torque_LBK}). This time-asymptotic limit not only ignores transient phenomena relevant in real galaxies, but also leads to qualitatively wrong conclusions: in the limit $t \rightarrow \infty$, LBK predicts that contribution from non-resonant orbits vanish by phase mixing and only the perfectly resonant orbits give rise to a \textit{non-zero} torque, whereas in truth, the resonant orbits would also phase mix within the trapped zone resulting in a \textit{zero} torque as shown in Fig.~\ref{fig:torque_W04}.

\begin{figure*}
  \begin{center}
      \includegraphics[width=17.0cm]{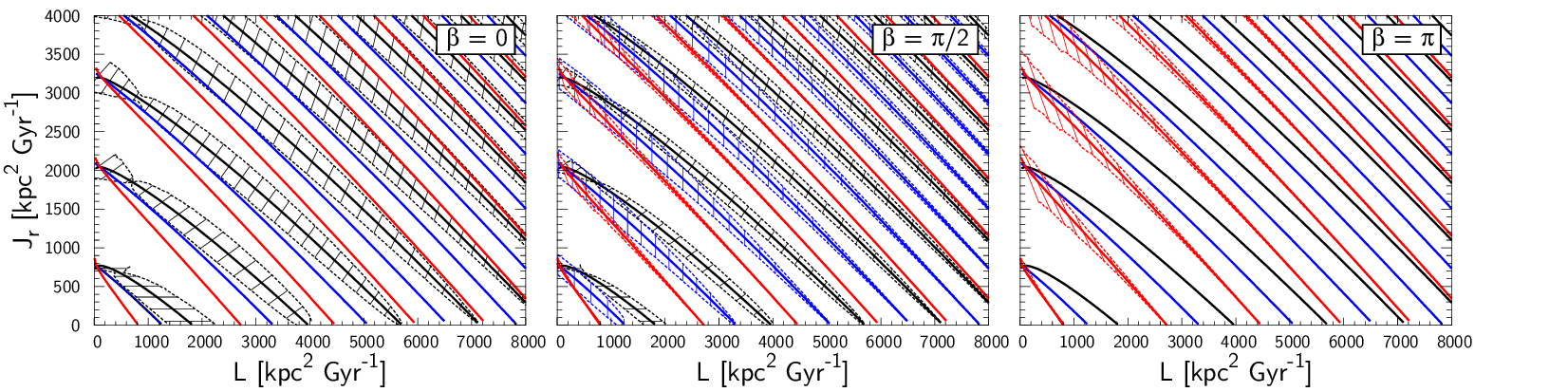}
    \vspace{-2mm}
    \caption{Overlap of resonances with $\Npsi = 2$ (black), $\Npsi = 0$ (blue) and $\Npsi = -2$ (red). See text for detail.}
    \label{fig:app_resonance_overlap_J}
  \end{center}
\end{figure*}

\section{Elliptic integrals}
\label{sec:app_elliptic_functions}

The complete elliptic integral of the first $K$ and second $E$ kinds are
\begin{align}
  K(k) = \int_0^{\frac{\pi}{2}} \frac{\drm \theta}{\sqrt{1 - k^2 \sin^2 \theta}},~~~
  E(k) = \int_0^{\frac{\pi}{2}} \drm \theta \sqrt{1 - k^2 \sin^2 \theta}.
  \label{eq:app_comp_ell_int}
\end{align}
The incomplete elliptic integral of the first kind $F(\phi|k)$ is
\begin{align}
  &u = F(\phi|k) \equiv \int_0^{\phi} \frac{\drm \theta}{\sqrt{1 - k^2 \sin^2 \theta}}.
  \label{eq:app_incomp_ell_int}
\end{align}
and the Jacobi elliptic functions are defined via the inverse of $F(\phi|k)$:
\begin{align}
  &\phi = F^{-1}(u|k), \\
  &{\rm sn}(u|k) \equiv \sin \phi,~~
  {\rm cn}(u|k) \equiv \cos \phi,~~
  {\rm dn}(u|k) \equiv \sqrt{1 - k^2 \sin^2 \phi}.
  \label{eq:app_Jacobi_ell_func}
\end{align}
The derivatives of the Jacobi elliptic functions are
\begin{align}
  \frac{\drm}{\drm u} {\rm sn}(u|k) &= {\rm cn}(u|k) {\rm dn}(u|k), ~~~
  \frac{\drm}{\drm u} {\rm cn}(u|k) = - {\rm sn}(u|k) {\rm dn}(u|k), \nonumber \\
  \frac{\drm}{\drm u} {\rm dn}(u|k) &= - k^2 {\rm sn}(u|k) {\rm cn}(u|k).
\label{eq:app_Jacobi_ell_func_derivatives}
\end{align}

\section{Overlap of resonances}
\label{sec:app_overlap_of_resonances}

Figure~\ref{fig:app_resonance_overlap_J} overlays all three sets of resonances with $\Npsi = 2$ (black), $\Npsi = 0$ (blue) and $\Npsi = -2$ (red). As in Fig.~\ref{fig:f_res_J}, $\Nphi$ is 2 for all resonances and $\Nr$ increase from bottom left to top right: 0 to 9 (black), 1 to 10 (blue), and 2 to 11 (red). The thick solid curves mark the resonance, the dotted curves mark the maximum width of the trapped zone, and the thin lines indicate the direction of libration. All three resonances intersect near $L=0$ where $\Omegar = 2\Omegapsi$ is satisfied. Overlap of bar resonances is absent at orbital inclination $\beta = 0$ (left) and $\beta = \pi$ (right), implying that chaos due to resonance overlap is of less concern in disc dynamics. At $\beta = \pi/2$ (middle), we see resonances overlapping particularly at low $L$ and high $\Jr$.

\section{Orbit subject to a slowing bar}
\label{sec:app_orbit_subject_to_slowing_bar}

We model an orbit trapped and dragged by a slowing bar using the Hamiltonian (\ref{eq:Hamiltonian}) with a time-dependent pattern speed $\Omegap(t)$,
\begin{align}
  \bH(\thetas,\vJ',t)
  &= H_0(\vJ') - \Nphi \Omegap(t) \Js + \Psi(\vJ') \cos \left(\thetas - \thetasres\right).
  \label{eq:Hamiltonian_wp_t}
\end{align}
As before, we Taylor expand around $\Js = \Jsres$ which satisfies the resonance condition at $t = \tres$:
\begin{align}
  \bH(\thetas,\vJ',t)
  &\simeq \left[H_0(\vJ') - \Nphi \Omegap(t) \Js\right]_{\Jsres} \! + \left[\frac{\pd H_0}{\pd \Js} - \Nphi \Omegap(t)\right]_{\Jsres} \!\! \Delta \nonumber\\
  &+ \frac{1}{2} G(\vJf,\Jsres) \Delta^2 + \Psi(\vJf,\Jsres) \cos \theta,
  \label{eq:app_Hamiltonian_wp_t_expand}
\end{align}
where we wrote $\Delta \equiv \Js - \Jsres$, $\theta \equiv \thetas - \thetasres$ and $G \equiv \frac{\pd^2 H_0}{\pd \Js^2}$ for convenience. Removing the first bracket term, which does not affect the slow dynamics, and substituting $\frac{\pd H_0}{\pd \Js}\bigr\vert_{\Jsres} = \vN \cdot \vOmega(\vJf,\Jsres) = \Nphi \Omegap(\tres)$ in the second bracket term yields \citep{weinberg1994kinematic}
\begin{align}
  \bH(\thetas,\Js,t)
  &= - \Nphi \left[\Omegap(t) - \Omegap(\tres)\right] \Delta + \frac{1}{2} G \Delta^2 + \Psi \cos \theta.
  \label{eq:app_Hamiltonian_wp_t_expand_2}
\end{align}
From the Hamilton's equations of motion:
\begin{align}
  \dtheta = - \Nphi \left[\Omegap(t) - \Omegap(\tres)\right] + G \Delta ~~{\rm and}~~
  \dot{\Delta} = \Psi \sin \theta,
  \label{eq:app_Hamiltons_equation_wp_t}
\end{align}
we obtain 
\begin{align}
  \ddtheta + \omega_0^2 \left(\sin \theta - s \right) = 0,
  \label{eq:app_ddtheta_s}
\end{align}
where $\omega_0^2 \equiv - G\Psi$, and $s = - \Nphi \dOmegap/\omega_0^2$ is the speed parameter. If we assume that $\dOmegap$ vary little during the typical libration period $2\pi / \omega_0$, $s$ can be assumed constant and we obtain the following energy integral:
\begin{align}
  \Ep = \frac{1}{2} {\dtheta}^2 + V(\theta) ~~{\rm where}~~ V(\theta) = \omega_0^2 \left( - \cos \theta - s \theta \right).
  \label{eq:app_Ep}
\end{align}
Trapped orbits are confined in the potential well of $V(\theta)$ which is linearly tilted due to the second term \citep[see Fig.4 of][]{Chiba2020ResonanceSweeping}. The crest of the potential is at $\thetasep = \sin^{-1}(s)~~ (\pi/2 \leq \thetasep \leq \pi)$ and the minimum energy $\Ep$ required to reach the crest is 
\begin{align}
  \Epsep = \omega_0^2 \left( - \cos \thetasep - s \thetasep \right).
  \label{eq:app_Epsep}
\end{align}
Orbit with $\Ep = \Epsep$ approximately marks the boundary of phase space that remains trapped in resonance. Hence, by a slight abuse of language, we will refer to the orbit $\Ep = \Epsep$ as the `separatrix'. Inserting (\ref{eq:app_Epsep}) and (\ref{eq:app_Hamiltons_equation_wp_t}) to (\ref{eq:app_Ep}) yields an equation for the separatrix:
%
\begin{align}
  \Delta_{\pm}
  =& \frac{\Nphi \left[\Omegap(t) - \Omegap(\tres)\right]}{G} \nonumber \\
  &\pm \frac{\omega_0}{G} \sqrt{2 \left[\cos\theta - \cos\thetasep + s \left(\theta - \thetasep \right)\right]}.
  \label{eq:app_Delta_theta_relation}
\end{align}
The first term describes the drift of the resonance and the second term describes the shape of the separatrix. Since the first term is time dependent, the separatrix is not closed in phase space. However, in the slow limit $s \ll 1$ the distance the resonance travels in a typical libration period is smaller than the size of the resonance, so we may mark the approximate area of trapped phase space by neglecting the first term. This is drawn in Fig.~\ref{fig:phase_flow_slowAA_slow}. In comparison to the separatrix of the time-frozen Hamiltonian (dot-dashed black), the separatrix of the time-dependent Hamiltonian (solid black) is contracted and shifted towards $\theta$, describing the trapped boundary of the simulation remarkably well. The first term in fact breaks/opens the separatrix and thus allows orbits to enter or escape the trapped region.

Setting $s=0$ in equation (\ref{eq:app_Delta_theta_relation}), we recover the equation for the separatrix of a standard pendulum Hamiltonian \citep[e.g.][]{lichtenberg1992regular}:
\begin{align}
  &\Delta_{\pm} = \pm \frac{2\omega_0}{G} \cos \frac{\theta}{2}.
  \label{eq:app_Delta_theta_relation_s0}
\end{align}
%


\bsp	
\label{lastpage}
\end{document}

%% file: newcommand
\def\eqref#1{equation~(\ref{#1})}

\newcommand {\Myr}{\,{\rm Myr}}
\newcommand {\Gyr}{\,{\rm Gyr}}
\newcommand {\iGyr}{\,{\rm Gyr}^{-1}}

\newcommand {\kpc}{\,{\rm kpc}}
\newcommand {\kms}{\,{\rm km}\,{\rm s}^{-1}}

\newcommand {\kpckpcGyr}{\,{\rm kpc}^2\,{\rm Gyr}^{-1}}
\newcommand {\kmskpc}{\,{\rm km}\,{\rm s}^{-1}\,{\rm kpc}^{-1}}

\newcommand {\Msun}{\,{\rm M}_\odot}

\newcommand {\vc}{v_{\rm c}}

\newcommand {\drm}{\mathrm{d}}

\newcommand {\vx}{{\bm x}}
\newcommand {\vvel}{{\bm v}}

\newcommand {\Lz}{L_z}

\newcommand {\phib}{\varphi_{\rm b}}

\newcommand {\rCR}{r_{\rm CR}}
\newcommand {\vJ}{{\bm J}}
\newcommand {\Jr}{J_r}

\newcommand {\vJf}{{\bm J}_{\rm f}}
\newcommand {\Jfo}{J_{{\rm f}_1}}
\newcommand {\Jft}{J_{{\rm f}_2}}
\newcommand {\Js}{J_{\rm s}}

\newcommand {\Jsres}{J_{\rm s, res}}

\newcommand {\Jl}{J_\ell}
\newcommand {\Jc}{J_{\rm c}}
\newcommand {\Jlc}{J_{\ell/{\rm c}}}

\newcommand {\vtheta}{{\bm \theta}}
\newcommand {\thetar}{\theta_r}

\newcommand {\thetaphi}{\theta_\varphi}
\newcommand {\thetapsi}{\theta_\psi}
\newcommand {\thetas}{\theta_{\rm s}}

\newcommand {\thetafo}{\theta_{{\rm f}_1}}
\newcommand {\thetaft}{\theta_{{\rm f}_2}}
\newcommand {\dtheta}{\dot{\theta}}
\newcommand {\ddtheta}{\ddot{\theta}}
\newcommand {\thetares}{\theta_{\rm res}}
\newcommand {\thetasres}{\theta_{\rm s,res}}
\newcommand {\thetasep}{\theta_{\rm sep}}

\newcommand {\thetal}{\theta_\ell}
\newcommand {\thetac}{\theta_{\rm c}}
\newcommand {\thetalc}{\theta_{\ell/{\rm c}}}

\newcommand {\vthetaf}{{\bm \theta}_{\rm f}}
\newcommand {\Omegap}{\Omega_{\rm p}}
\newcommand {\Omegapo}{\Omega_{\rm p0}}

\newcommand {\dOmegap}{\dot{\Omega}_{\rm p}}

\newcommand {\Omegar}{\Omega_r}

\newcommand {\Omegapsi}{\Omega_\psi}
\newcommand {\vOmega}{\mathbf \Omega}

\newcommand {\Omegafo}{\Omega_{{\rm f}_1}}
\newcommand {\Omegaft}{\Omega_{{\rm f}_2}}
\newcommand {\Omegas}{\Omega_s}
\newcommand {\Omegal}{\Omega_\ell}
\newcommand {\Omegac}{\Omega_{\rm c}}
\newcommand {\Omegalc}{\Omega_{\ell/{\rm c}}}

\newcommand {\vn}{{\mathbf n}}
\newcommand {\vN}{{\bm N}}
\newcommand {\vk}{{\bm k}}
\newcommand {\vkf}{{\bm k}_{\rm f}}
\newcommand {\vm}{{\mathbf m}}

\newcommand {\nr}{{\rm n}_r}
\newcommand {\npsi}{{\rm n}_\psi}
\newcommand {\nphi}{{\rm n}_\varphi}

\newcommand {\Nr}{N_r}
\newcommand {\Npsi}{N_\psi}
\newcommand {\Nphi}{N_\varphi}
\newcommand {\ks}{k_{\rm s}}

\newcommand {\kfo}{k_{{\rm f}_1}}
\newcommand {\kft}{k_{{\rm f}_2}}

\newcommand {\Phib}{\Phi_{\rm b}}

\newcommand {\Philm}{\Phi_{lm}}

\newcommand {\hPhi}{\hat{\Phi}}
\newcommand {\hPsi}{\hat{\Psi}}

\newcommand {\hPhin}{\hat{\Phi}_{\vn}}

\newcommand {\hPsik}{\hat{\Psi}_{\vk}}
\newcommand {\hPsim}{\hat{\Psi}_{\vm}}

\newcommand {\hf}{\hat{f}}

\newcommand {\hfn}{\hat{f}_{\vn}}

\newcommand {\e}{\mathrm{e}}
\newcommand {\bH}{\bar{H}}
\newcommand {\Ep}{E_{\rm p}}
\newcommand {\Epsep}{E_{\rm {p,sep}}}

\newcommand {\tres}{t_{\rm res}}

\newcommand {\Tl}{T_\ell}

\newcommand {\rs}{r_{\rm s}}
\newcommand {\vg}{v_g}

\newcommand {\pd}{{\partial}}

\newcommand {\torq}{\left\langle\frac{\drm \Lz}{\drm t}\right\rangle}

\newcommand {\Ib}{I_{\rm b}}

\newcommand {\varE}{\mathcal{E}}